\documentclass[acmsmall]{acmart}

\AtBeginDocument{%
  \providecommand\BibTeX{{%
    \normalfont B\kern-0.5em{\scshape i\kern-0.25em b}\kern-0.8em\TeX}}}

\setcopyright{acmlicensed}
\acmJournal{PACMHCI}
\acmYear{2021} 
\acmVolume{5} 
\acmNumber{CSCW2} 
\acmArticle{418} 
\acmMonth{10} 
\acmPrice{15.00} 
\acmDOI{10.1145/3479562}

\begin{document}

\title[Algorithmic Risk Assessments Can Alter Human Decision-Making Processes]{Algorithmic Risk Assessments Can Alter Human Decision-Making Processes in High-Stakes Government Contexts}

\author{Ben Green}
\affiliation{%
  \institution{University of Michigan}
  \city{Ann Arbor}
  \state{MI}
  \country{USA}}
\email{bzgreen@umich.edu}

\author{Yiling Chen}
\affiliation{%
  \institution{Harvard University}
  \city{Cambridge}
  \state{MA}
  \country{USA}}
\email{yiling@seas.harvard.edu}


\begin{abstract}
Governments are increasingly turning to algorithmic risk assessments when making important decisions, such as whether to release criminal defendants before trial. 
Policymakers assert that providing public servants with algorithmic advice will improve human risk predictions and thereby lead to better (e.g., fairer) decisions.
Yet because many policy decisions require balancing risk-reduction with competing goals, improving the accuracy of predictions may not necessarily improve the quality of decisions.
If risk assessments make people more attentive to reducing risk at the expense of other values, these algorithms would diminish the implementation of public policy even as they lead to more accurate predictions. 
Through an experiment with 2,140 lay participants simulating two high-stakes government contexts, we provide the first direct evidence that risk assessments can systematically alter how people factor risk into their decisions.
These shifts counteracted the potential benefits of improved prediction accuracy.
In the pretrial setting of our experiment, the risk assessment made participants more sensitive to increases in perceived risk; this shift increased the racial disparity in pretrial detention by 1.9\%. 
In the government loans setting of our experiment, the risk assessment made participants more risk-averse; this shift reduced government aid by 8.3\%. 
These results demonstrate the potential limits and harms of attempts to improve public policy by incorporating predictive algorithms into multifaceted policy decisions.
If these observed behaviors occur in practice, presenting risk assessments to public servants would generate unexpected and unjust shifts in public policy without being subject to democratic deliberation or oversight.
\end{abstract}



\maketitle

\section{Introduction}
Following recent advances in the quality and accessibility of algorithms, governments increasingly use machine learning when making high-stakes decisions \cite{green2019sec, eubanks2018automating}. 
Many applications of algorithms involve risk assessments, which predict the risk of some adverse outcome.
These predictions are then presented to human decision-makers to inform consequential decisions about individuals. 
Applications of public sector risk assessments include informing pretrial and sentencing decisions with a criminal defendant's likelihood to recidivate \cite{loomis2016, nj2017report}, targeting public health inspections based on a child's risk of lead poisoning \cite{potash2015lead}, and directing child welfare interventions based on a child's risk of being abused or neglected \cite{eubanks2018automating}.

Although machine learning algorithms are adopted and celebrated for their accuracy in predicting policy outcomes \cite{kleinberg2018predictions, kleinberg2015policy}, in practice algorithms typically assist people in making informed yet ultimately normative decisions. 
Risk predictions and policy decisions represent distinct tasks: unlike predictions of risk, which can be directly optimized for accuracy, many policy decisions require balancing competing goals and therefore lack a straightforward correct answer \cite{zacka2017state}. 

The particular balancing act for any policy decision is of normative significance and is often subject to vigorous debate.
How frontline government officials weigh competing values when making decisions effectively determines the implementation of public policy \cite{lipsky1980slb}.
This makes it imperative that government decision-makers strike the appropriate balance between the conflicting goals embedded in public policy \cite{zacka2017state}. 
In settings such as pretrial detention, ``[h]ow this balance is struck [\ldots] has enormous implications'' \cite{mahoney2001pretrial}.

The distinction between predictions and decisions means that using risk assessments to improve people's predictions may not necessarily improve people's decisions.
In particular, the normative multidimensionality inherent in many government decisions can create conflicts between risk-reduction and other values.
For instance, although pretrial decisions must limit the risk of defendants being rearrested or not returning to court for trial, they must also prioritize the liberty of defendants \cite{aba2007pretrial}.
Similarly, although government loan decisions must limit the risk of recipients defaulting on their loans, they must also promote equity by supporting low-income applicants \cite{usda2020single}.

Thoroughly evaluating the impacts of risk assessments therefore requires considering not just whether these algorithms improve the accuracy of human \emph{predictions}, but also whether they improve the quality of human \emph{decisions}.
One notable concern regarding risk assessments is that these tools could alter how people balance risk with other considerations when making decisions \cite{green2018fatml, starr2014sentencing}.
Although improving the accuracy of risk predictions is appropriate for policies that include risk as a consideration, any systematic change in the salience of risk in decision-making would amount to a shift in public policy, with potentially disparate impacts. 
As a result, more than 100 civil rights and social justice organizations have raised concerns that the emphasis on risk by pretrial risk assessments will prompt judges to treat defendants more harshly \cite{leadership2018pretrial}.

Policymakers and others advocating for risk assessments assert that these algorithms merely improve human predictions without altering how people factor risk into their decisions \cite{eubanks2018automating, malenchik2010, nj2017report, starr2014sentencing, loomis2016}. 
Proponents claim that as long as decision-makers are granted autonomy and discretion over final decisions, providing public servants with accurate risk predictions will lead these individuals to make better (e.g., fairer and less punitive) decisions \cite{eubanks2018automating, malenchik2010, nj2017report, starr2014sentencing, loomis2016}. 
Yet despite being central to arguments for adopting public sector algorithms, this assumption has not been rigorously tested. 
In fact, this claim is called into question by mounting evidence that the implementation of algorithms often relies on untested---and false---assumptions about human-algorithm collaborations, leading to unjust outcomes \cite{green2021slate}.
For instance, criminal justice risk assessments have failed to generate the intended benefits because judges use these tools in unexpected ways \cite{stevenson2018assessing, stevenson2019hands, albright2019judge, brayne2020technologies}.

This paper tests the assumption that improving human prediction accuracy with algorithms will necessarily improve human decisions. 
In particular, we study how presenting a risk assessment's advice influences the decision-making processes of laypeople. 
We ran an online experiment with 2,140 U.S.-based participants recruited from Amazon Mechanical Turk (a widely used online platform for human subjects research \cite{coppock2019turk, komarov2013crowdsourcing}).
We used this experiment to explore the influence of risk assessments in two high-stakes government settings where decision-making involves balancing risk-reduction with other factors: a pretrial setting and a home improvement loans setting. 

We had three central goals for the study: 
1) determine whether risk assessments merely improve human risk predictions, as is commonly asserted, or also alter how people weigh risk in the decision-making process itself; 
2) characterize the effects of risk assessments on decision-making processes; 
and 3) determine how these effects impact outcomes such as racial disparities.
We hypothesized that presenting risk assessments would alter decision-making processes, prompting participants to become more attentive to avoiding risk when making decisions. 
We also hypothesized that this effect would exacerbate racial disparities in decisions. 

We found that risk assessments altered human decision-making processes in both settings.
Presenting risk assessments to participants systematically changed how they factored risk into policy-relevant decisions in ways that can lead to harmful outcomes. 
Although the risk assessments improved the accuracy of human predictions, they also induced shifts in human decision-making processes that counteracted the potential benefits of these enhanced predictions.
In the pretrial setting, the risk assessment made participants more sensitive to increases in perceived risk, increasing the racial disparity in pretrial detention by 1.9\%. 
In the loans setting, the risk assessment made participants more risk-averse at all levels of perceived risk, reducing government aid by 8.3\%.

These results challenge a central assumption behind support for algorithmic decision-making aids in government.
They demonstrate that improving human prediction accuracy with risk assessments does not necessarily improve human decisions and instead can have unexpected adverse consequences.
If these observed behaviors arose in practice, presenting algorithms to government decision-makers would generate unintended and unjust shifts in the application of public policy without being subject to democratic deliberation or oversight. 

Because we study laypeople in a lab setting, our results do not reflect the behaviors of experts making real decisions.
Practitioners differ from laypeople in numerous ways, as their specialized knowledge and professional identity shape their responses to risk assessments \cite{brayne2020technologies}.
However, there are several reasons to believe that our results could align with real-world outcomes and complement studies of expert decision-making in practice. 
First, research has found that both judges and financial professionals exhibit many of the same behaviors as laypeople. 
Judges are susceptible to cognitive and racial biases when making decisions in much the same manner as laypeople \cite{guthrie2001inside, rachlinski2008judges, rachlinski2018framing}. 
Similarly, financial professionals are susceptible to priming effects and loss aversion \cite{cohn2015evidence, gilad2008priming, haigh2005professional}.
Financial professionals exhibit these behaviors to a greater extent than laypeople \cite{gilad2008priming, haigh2005professional}; furthermore, greater professional experience among financial professionals does not mitigate priming effects \cite{cohn2015evidence}.
Second, experimental studies using procedures very similar to those in this study \cite{green2019fat, green2019cscw} have found behaviors among laypeople using risk assessments that align closely with behaviors observed among judges using risk assessments in practice \cite{albright2019judge, cowgill2018impact}. 

Experimental trials with laypeople therefore present a promising approach for evaluating and improving proposed human-algorithm collaborations before algorithmic decision-making aids are adopted. 
Performing these experimental studies would provide diagnostic knowledge that can inform experimental studies with experts and, in turn, the development, implementation, and evaluation of real-world systems.
Although the gold standard is data on how experts use risk assessments in practice, such evidence relies on retrospective data about systems that have been in use for years \cite{stevenson2018assessing, stevenson2019hands, hrw2017california, brayne2020technologies}.
By the time breakdowns in real-world human-algorithm collaborations are exposed, many people will have already been affected. 
Although some adverse behaviors would arise only in practice, many could potentially be detected before an algorithm's adoption via preliminary lab studies. 
Experimental trials can serve as an integral component of a broader evaluation pipeline, facilitating more rigorous and proactive scrutiny of whether algorithms actually improve decision-making.

\section{Background and Related Work}
The increasing use of algorithmic decision-making aids across government has placed novel human-algorithm collaborations at the center of consequential policy decisions.
In the context of machine learning, the standard framework for human-algorithm collaborations is ``human-in-the-loop'' systems. 
In these settings, people are incorporated into the machine learning pipeline to produce the best possible algorithmic output. 
Algorithms make the final decisions, with humans assisting (e.g., labeling training data and reviewing low-confidence classifications) \cite{appen2018hitl}.
In some cases, such tasks are completed by crowds of people, with crowdsourcing techniques used to support machine learning models \cite{wang2019crowd, kamar2012combining}.

In public policy contexts, however, human-algorithm collaborations involve a different process and goal: algorithms are incorporated into human decision-making processes to generate the best possible human decision. 
People make the final decisions, with algorithms assisting (e.g., making accurate predictions based on patterns within datasets). 
Instead of the human-in-the-loop frame, therefore, most uses of algorithms in government call for a complementary paradigm: ``algorithm-in-the-loop'' decision-making \cite{green2019fat}.
Algorithm-in-the-loop settings center human decisions---rather than algorithmic decisions---as the most important outcome, orienting attention to how algorithms influence human decisions. 

Despite the potential promise of algorithms aiding human predictions, experimental studies have uncovered numerous limits in people's ability to make appropriate and effective use of algorithmic advice. 
Several studies have found that algorithmic advice can improve the accuracy of human predictions, but people's decisions about when and how to diverge from algorithmic recommendations are typically incorrect \cite{green2019fat, green2019cscw, grgic2019human, lai2019deception}. 
People struggle to evaluate the quality of algorithmic advice \cite{goodwin1999forecasts, green2019fat, green2019cscw, lai2019deception}, often discount accurate algorithmic recommendations \cite{dietvorst2015aversion, lim1995forecasts, yeomans2017recommendations}, and exhibit racial biases in their responses to risk assessments \cite{green2019fat, green2019cscw}.
Although evidence suggests that experts are capable of overriding some erroneous predictions in practice \cite{dearteaga2020erroneous}, other evidence demonstrates that incorrect predictions reduce the quality of expert judgments \cite{kiani2020cancer} and that experts make less effective use of algorithmic forecasts than laypeople \cite{logg2019algorithm}.

These breakdowns in human-algorithm collaboration demonstrate that algorithmic interventions are indeterminate: the effects of using an algorithm often diverge in practice from what was expected based on the algorithm's technical characteristics \cite{green2020realism}.
Numerous evaluations of how judges use pretrial risk assessments indicate that providing accurate risk predictions does not generate the intended improvements in decision-making (i.e., reduced pretrial detention, recidivism, and racial disparities).
In jurisdictions across the country, judges disproportionately override release recommendations to detain defendants, leading to much higher than expected pretrial detention rates \cite{hrw2017california, sji2016report, steinhart2006aecf, stevenson2018assessing, stevenson2019hands}.
Several studies have shown that risk assessments exacerbate rather than diminish racial disparities in pretrial detention, in part because judges often make more punitive decisions about Black defendants than similar white defendants \cite{albright2019judge, cowgill2018impact, stevenson2018assessing}.
Furthermore, ethnographic work has found that judges often resist using risk assessments because they dislike the idea of these tools replacing or surveilling them \cite{brayne2020technologies}. 

Despite growing knowledge about how people use algorithms in both experimental and real-world settings, a significant open question is whether presenting algorithmic advice alters the process through which people make decisions.
Because risk assessments emphasize the likelihood of specific adverse outcomes (such as a pretrial defendant failing to appear for trial or being rearrested), scholars and activists have raised concerns that risk assessments could make risk a more salient factor in decision-making processes \cite{green2018fatml, starr2014sentencing, leadership2018pretrial}. 
Such concerns follow closely from the existing literature on framing and priming effects. 
Prior research demonstrates that framing decisions around losses motivates decision-makers (including judges) to avoid those losses \cite{rachlinski2018framing, scurich2011framing, tversky1981framing}. 
Similarly, priming people (including financial professionals) to consider risks makes them less likely to make or support decisions that involve risk \cite{cohn2015evidence, eckles2011priming, erb2002choice, gilad2008priming}.

Initial studies have observed behaviors that are consistent with the possibility of risk assessments altering decision-making processes. 
A small experiment with 83 law students making simulated sentencing decisions found that presenting a risk assessment increased the sentence given to a high-risk defendant and decreased the sentence given to a low-risk defendant \cite{starr2014sentencing}. 
A later experiment with 340 judges making simulated sentencing decisions found that presenting a risk assessment increased the likelihood that a low socioeconomic status defendant would be incarcerated and decreased the likelihood that a high socioeconomic status defendant would be incarcerated \cite{skeem2019impact}. 

Although these results suggest that showing a risk assessment heightens the salience of risk, they cannot conclusively determine whether risk assessments alter how people weigh risk when making decisions.
Because these studies looked only at people's decisions with and without a risk assessment, neither can distinguish between two potential explanations: a) the risk assessment actually increased the weight that people gave to reducing risk when making decisions, and b) the risk assessment merely influenced the estimates of risk that people factored into their decisions.
Distinguishing between these two explanations requires accounting for a risk assessment's effects on human risk predictions rather than directly comparing human decisions with and without a risk assessment.

\section{Hypotheses and Framework}
In this study, we investigate whether risk assessments systematically alter how people factor risk into their decisions. 
Because risk assessments emphasize the risk of an adverse outcome, we hypothesized that presenting risk assessments would make people more attentive to avoiding risk when making decisions.
Given existing racial disparities in risk, we also hypothesized that this effect would exacerbate racial disparities in decisions. 

\begin{figure}
	\centering
	\includegraphics[width=\columnwidth]{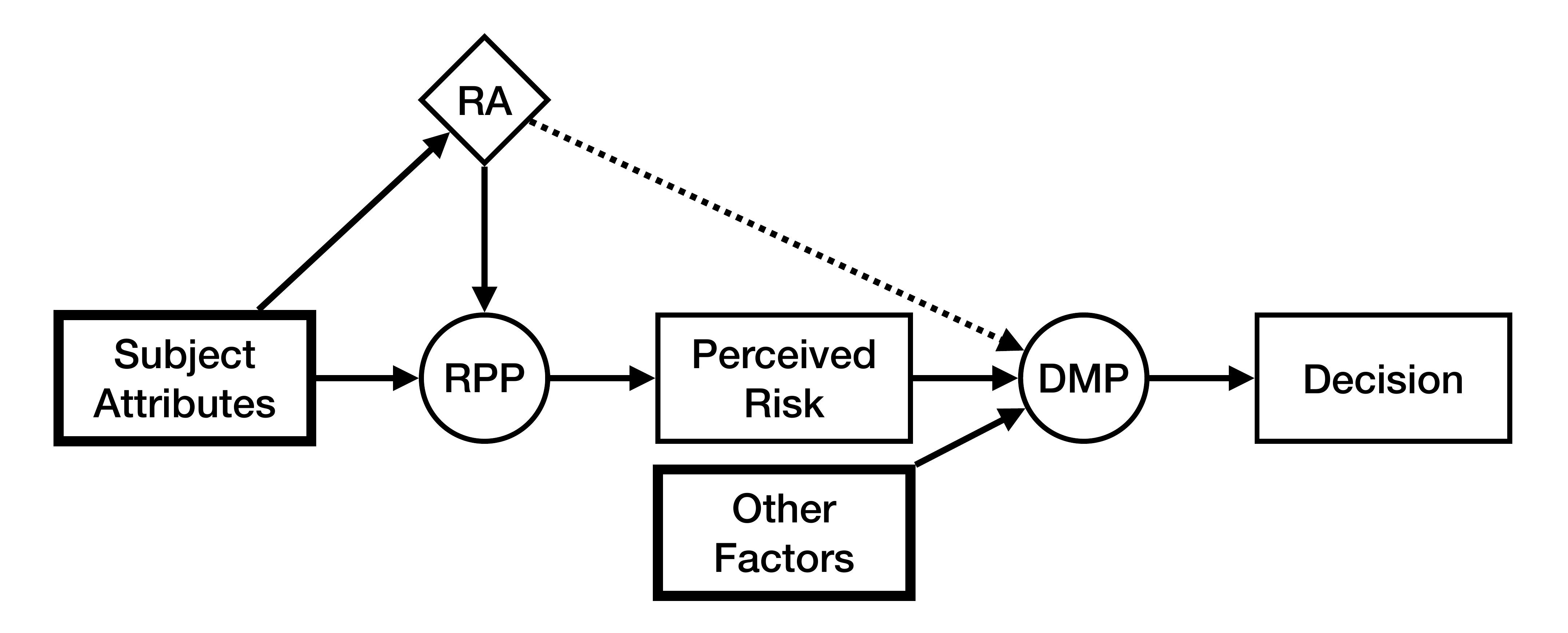}
	\Description{A diagram representing how law and policy conceptualize the risk-prediction process (RPP) and the decision-making process (DMP) when people interact with risk assessments.}
  	\caption{How subject attributes are translated into a decision with the aid of a risk assessment, as conceptualized in law and policy. 
Circles represent the two stages of human cognitive processing: the risk-prediction process (RPP) and the decision-making process (DMP). 
The dashed line from the risk assessment (RA) to the DMP represents the key question of this study: whether the RA alters the DMP. 
The absence of this influence (i.e., the absence of the dashed line) represents Scenario 3.
The presence of this influence (i.e., the presence of the dashed line) represents Scenario 4. 
Bold lines indicate that the rectangle represents a set of multiple attributes or factors.}
  	\label{fig:dmp}
 \end{figure}

Based on how policy documents \cite{nj2017report} and court rulings \cite{malenchik2010, loomis2016} describe the use of risk assessments, we analyzed decisions as being made through a two-stage process (Figure~\ref{fig:dmp}). 
First is the risk-prediction process (RPP), which represents a quantitative prediction task.
The RPP evaluates the attributes of a given subject (e.g., pretrial defendant) to predict that person's risk of an adverse outcome (e.g., failing to return to court for trial or being arrested before trial). 
This process yields a decision-maker's ``perceived risk'' about the subject.
Second is the decision-making process (DMP), which involves a normative balancing act between numerous considerations rather than a straightforward translation of risk into a decision.
The DMP incorporates the perceived risk alongside other relevant factors (e.g., the harms associated with pretrial detention) to make a decision about the subject (e.g., whether to release the defendant before their trial). 
A systematic change to the DMP reflects a shift in how decision-makers balance risk with other factors, which amounts to a shift in public policy \cite{mahoney2001pretrial, zacka2017state}. 
We instantiated the DMP as a function that determines the probability of detaining a defendant or rejecting a loan applicant conditioned on the perceived risk about the subject in question.\footnote{From a decision-making standpoint, what matters as an input to the DMP is the decision-maker's perceived risk, as that is what decision-makers act on. The perceived risk can be influenced by the risk assessment, among other factors.}

\begin{table}
\caption{The four possible ``scenarios'' of how a risk assessment (RA) can affect the risk-prediction process (RPP) and the decision-making process (DMP). 
Scenario 1 represents a baseline process without a risk assessment. 
Scenarios 2--4 represent the possible outcomes when decision-makers are presented with a risk assessment (Scenario 2 is ruled out by prior research, however). 
}
\begin{tabular}{p{1.5in} p{1.6in} p{1.6in}}
\toprule
 & \textbf{DMP unaffected by RA} & \textbf{DMP affected by RA} \\ \midrule
\textbf{RPP unaffected by RA} & Scenario 1 (Baseline: RA does not affect RPP or DMP) & Scenario 2 (Implausible: RA affects only DMP) \\
\textbf{RPP affected by RA} & Scenario 3 (Common assumption: RA affects only RPP) & Scenario 4 (Hypothesis: RA affects both RPP and DMP) \\ \bottomrule
\end{tabular}
\label{tab:scenarios}
\end{table}

The central question of this study is whether risk assessments influence only the RPP, as is typically assumed, or instead affect both the RPP \emph{and} the DMP. 
We categorize the influence of risk assessments into four possible ``scenarios,'' as summarized in Table~\ref{tab:scenarios}. 
Scenario 1 represents the baseline condition without any risk assessment. 
When a risk assessment's advice is presented, it could lead to either Scenario 3 or Scenario 4.\footnote{Scenario 2, in which risk assessments alter the DMP but not the RPP, is ruled out by prior research demonstrating that risk assessments influence human predictions \cite{green2019fat, green2019cscw, grgic2019human, stevenson2019hands}.} 
Scenario 3 is the commonly assumed outcome: risk assessments alter the RPP but not the DMP, meaning that improvements in prediction accuracy lead directly to more informed decisions.
The assumption that algorithms lead to Scenario 3 is central to support for algorithmic decision-making aids in government \cite{eubanks2018automating, malenchik2010, nj2017report, starr2014sentencing, loomis2016}. 
In this scenario, which represents the absence of the dashed line in Figure~\ref{fig:dmp}, improving prediction accuracy with risk assessments directly improves decisions. 
Scenario 4 is our hypothesized outcome: risk assessments alter both the RPP and the DMP, meaning that shifts in the DMP could counteract any gains in prediction accuracy.
In this scenario, which represents the presence of the dashed line in Figure~\ref{fig:dmp}, improving prediction accuracy with risk assessments may not improve decisions. 

Our goal is to test whether showing a risk assessment alters the DMP, which amounts to distinguishing between Scenario 3 and Scenario 4.
However, we cannot directly observe the DMP because it is a latent form of cognitive processing. 
Because risk assessments alter the RPP \cite{green2019fat, green2019cscw, grgic2019human, stevenson2019hands}, simply showing that a risk assessment changed participant decisions is insufficient to demonstrate that the risk assessment affected the DMP.
The challenge arises because a risk assessment could alter decisions in two distinct but superficially indistinguishable ways.
First, it could influence the RPP alone (Scenario 3), which would lead to decisions based on different risk estimates without changing how perceived risk factors into decisions. 
Second, it could influence both the RPP and the DMP (Scenario 4), which would lead to decisions based on different risk estimates \emph{and} would change how perceived risk factors into decisions. 

The only way to determine whether risk assessments affect the DMP is to compare decisions made with and without a risk assessment \emph{while accounting for the risk assessment's effects on predictions}. 
Accomplishing this requires access to decision-makers' perceptions of risk about each subject.
However, this information is not produced in practice and is difficult to obtain experimentally without influencing people's behavior. 
Determining the risk assessment's effects on the DMP thus requires a more complex experimental setup than prior work that has directly compared the decisions that people make with and without a risk assessment's advice. 
As described in more detail below, we designed our experiment to elicit both decisions and predictions from participants, enabling us to infer the effects of risk assessments on human decision-making processes.

\section{Methods}
Our study progressed in two stages. 
The first stage involved developing risk assessments for pretrial detention and home improvement loans. 
The second stage involved running an experiment on Amazon Mechanical Turk to evaluate how people interact with these risk assessments when making predictions and decisions. 
The full study was approved by the Harvard University Institutional Review Board and the National Archive of Criminal Justice Data (which manages the data used for the pretrial setting). 

\subsection{Study Settings}
Our experiment simulated two settings of government decision-making: pretrial detention and government home improvement loans.
Within the context of this study, these settings are structurally similar. 
In both settings, decision-makers must balance risk-reduction with conflicting normative considerations. 
Decisions are made about an individual ``subject'' and can be ``positive decisions'' or ``negative decisions.''
Subjects with high risk are more likely to receive negative decisions.
In the pretrial setting, the subject of the decision is a criminal defendant, the positive decision is to release the defendant before trial, and the negative decision is to detain the defendant before trial. 
In the loans setting, the subject of the decision is a loan applicant, the positive decision is to approve the loan application, and the negative decision is to reject the loan application.

\subsubsection{Pretrial Detention Setting}
After someone is arrested in the United States, they must await trial. 
Courts can either hold the criminal defendant in jail until their trial or release them with a mandate to return for their trial.\footnote{In practice, many defendants are also released under conditions such as paying a cash bond or being subject to electronic monitoring.} 
Pretrial detention decisions involve balancing competing goals. 
Courts aim to ensure that defendants will return to court for trial and will not commit any crimes if released. 
The higher the risk that a defendant will fail to return to court for their trial or will commit any crimes, the more likely a judge is to detain the defendant until their trial.
The interest in reducing risk is enhanced by detaining defendants. 
However, pretrial decisions are also made with an interest in protecting the liberty of defendants, ensuring that defendants are able to mount a proper legal defense, and reducing the hardship to defendants and their families \cite{aba2007pretrial}. 
Pretrial detention is associated with a range of negative outcomes that include longer prison sentences, sexual abuse, and limited employment opportunities \cite{green2020risk}. 
The interests in protecting liberty and avoiding the harms of pretrial detention are advanced by releasing defendants. 

In recent years, many jurisdictions across the U.S. have turned to risk assessments as a tool to make more accurate and objective predictions of risk. 
These improvements in prediction are intended to reduce racial biases and increase pretrial release rates \cite{harris2017pretrial,jensen2012hb463,nj2017report}.

\subsubsection{Government Home Improvement Loans Setting}
Many people apply for a loan to improve their house (e.g., to rehabilitate a home or to make a home energy efficient). 
When someone applies for a loan, it is common for the lender to assess the risk that the borrower will fail to pay back the money. 
This is known as defaulting on the loan. 
The higher the risk that the potential borrower will default on the loan, the less likely the lender generally is to provide money to that person. 
The U.S. government provides many types of home improvement loans in order to support low-income applicants who are unable to obtain affordable loans from banks \cite{usda2020single}. 
This sets up a balancing act between conflicting aims.
On the one hand, the goal of limiting loan default risk is enhanced by declining loans to low-income applicants. 
On the other hand, the goals of promoting equity, economic development, and community stability are enhanced by providing loans to low-income applicants. 

It is common for lenders to evaluate loan applicants using risk assessments that predict the likelihood of loan default.
Although there are no known cases of governments using risk assessments when allocating home improvement loans, this setting is akin to government uses of risk assessments to determine who should receive other resources \cite{eubanks2018automating}. 

\subsection{Risk Assessments}
In order to test the effects of presenting risk assessment predictions to participants in our experiment, we first developed risk assessments for pretrial detention and government home improvement loans. 
See Section~\ref{app:RAs} of the Appendix for a more detailed description of the data we used and how we developed these models. 
Our goal in this stage was not to develop optimal risk assessments, but to develop risk assessments that resemble those used in practice and that could be presented to participants during the Mechanical Turk experiment. 
We used datasets with information about 47,141 felony defendants across the United States who had been released before trial \cite{doj2009icpsr} (Table~\ref{tab:defendants}) and 45,218 recipients of home improvement loans via the peer-to-peer lending company Lending Club (Table~\ref{tab:loans}). 
The data included demographic information (including race, which we restricted to Black and white) for the felony defendants but not the loan applicants. 

We developed risk assessments (i.e., machine learning classifiers) using gradient boosted trees with ten-fold cross-validation. 
Our models included five attributes of each defendant\footnote{Defendant factors: age, offense type, number of prior arrests, number of prior convictions, and whether that person has any prior failures to appear for trial.} and seven attributes of each loan application.\footnote{Loan application factors: the applicant's annual income, credit score, and home ownership, as well as the loan's value, interest rate, monthly installment, and term of repayment.} 
The pretrial risk assessment was trained to predict whether a defendant, if released before trial, would fail to appear in court for trial or would be arrested before trial. 
The loans risk assessment was trained to predict whether a loan applicant, if given the loan, would default on that loan. 
Both risk assessments exhibited similar accuracy to pretrial and loan risk assessments developed in research and practice (pretrial AUC=0.67, loans AUC=0.69). 
Drawing from the held-out validation sets in each setting, we selected samples of 300 defendants and 300 loan applicants whose profiles and risk predictions would be presented to participants during the experiment. 
When used in our experiment, the risk assessments presented numerical predictions of risk about subjects (i.e., 0\%--100\%, in intervals of 10\%) but did not suggest what decision participants should make based on those predictions. 

\subsection{Experimental Design}
We recruited 2,685 participants on Amazon Mechanical Turk over two weeks in May 2020, restricting our task to workers inside the United States who had a task approval rate of at least 75\%.\footnote{In light of the COVID-19 pandemic, immediately before running this experiment we replicated a trial experiment conducted in December 2019. Our tests demonstrated that the pandemic did not influence any of our results (see Section~\ref{app:covid} in the Appendix).}
Our analysis includes the results from the 2,140 participants who completed the experiment while also passing our quality control reviews (by correctly answering several comprehension questions and two attention-check questions). 
Across both settings, a majority of participants were male, white, and college graduates (Table~\ref{tab:participants}). 
Participants were paid \$3 for completing the experiment, and those making predictions received an additional payment of up to \$1 based on the accuracy of their predictions.
Bonus payments were allocated using a Brier score, which incentivizes participants to report their true estimate of risk.
Participants completed the experiment in an average of 19.0 minutes and received an average wage of \$15.02 per hour.

\begin{table}
\caption{Attributes of the participants in our experiment, by setting. Measures of familiarity, clarity, and enjoyment are based on participant self-reports measured on a Likert scale from 1 (low) to 7 (high).}
\begin{tabular}{@{}lll@{}}
\toprule
 & \textbf{Pretrial} & \textbf{Loans} \\ 
 & N=1,040 & N=1,100 \\ \midrule
\textbf{Demographics} &  &  \\
Male & 59.8\% & 61.0\% \\
Black & 14.2\% & 11.9\% \\
White & 71.5\% & 72.9\% \\
18-24 years old & 7.4\% & 6.8\% \\
25-34 years old & 46.1\% & 45.0\% \\
35-59 years old & 43.0\% & 43.9\% \\
60+ years old & 3.6\% & 4.3\% \\
College degree or higher & 82.5\% & 81.9\% \\
Criminal justice familiarity & 5.1 & 5.1 \\
Financial lending familiarity & 4.9 & 5.1 \\
Machine learning familiarity & 4.7 & 4.8 \\
 &  &  \\
\textbf{Outcomes} &  &  \\
Average experiment time & 19.1 minutes & 19.0 minutes \\
Average hourly wage & \$14.86 & \$15.16 \\
Experiment clarity & 6.4 & 6.4 \\
Participant enjoyment & 5.8 & 5.9 \\ \bottomrule
\end{tabular}
\label{tab:participants}
\end{table}

\begin{figure}
	\centering
	\includegraphics[width=\columnwidth]{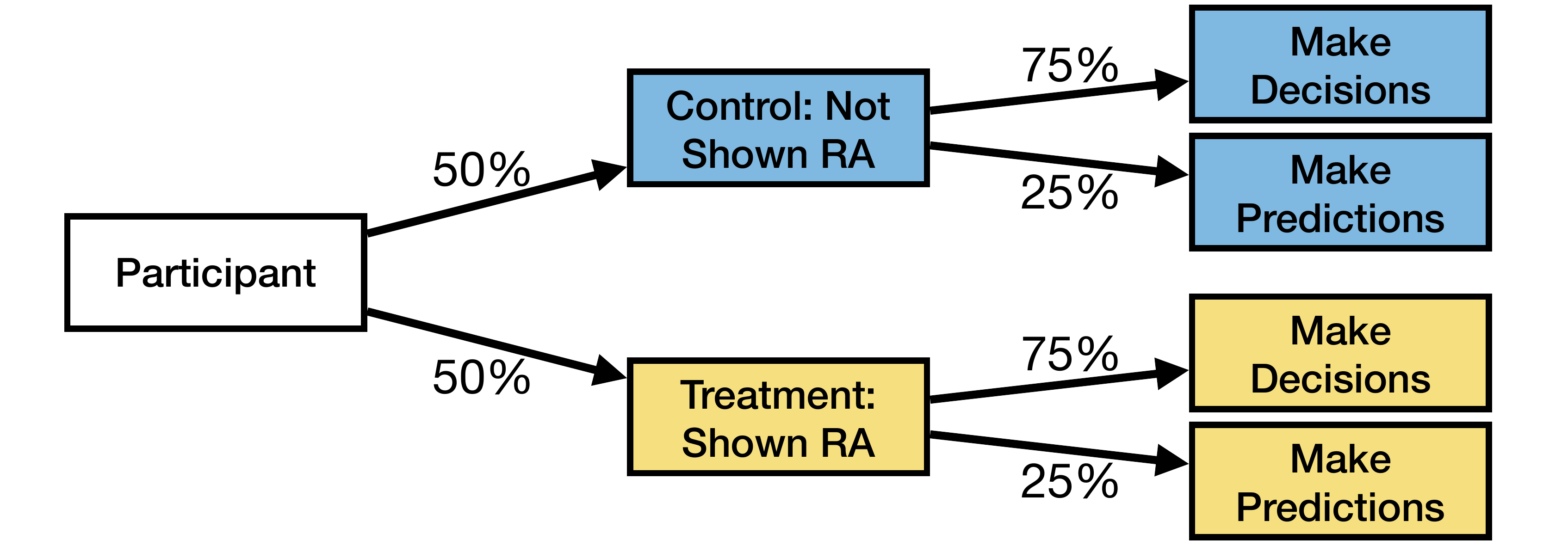}
	\Description{This diagram describes how participants were sorted into different conditions in each setting. Participants were first split into the control and treatment groups, with a 50\% probability to be in each. Participants were then split based on whether they would be asked to make decisions or predictions. There was 75\% chance of being the decisions group and 25\% chance of being in the predictions group.}
  	\caption{Our 2x2 experimental design and the four conditions that participants were sorted into in each setting. Probabilities indicate the likelihood of each path. Within each setting, participants in all four conditions were presented with a sample of subjects drawn from the same set of 300 subjects.}
  	\label{fig:conditions}
 \end{figure}

The experimental design consisted of three treatments. 
When participants entered the experiment, they were split evenly into either the pretrial or loans setting. 
We then followed a 2x2 design within each setting (Figure~\ref{fig:conditions}).
Participants were split into control groups (which were not presented with a risk assessment's advice) and treatment groups (which were presented with a risk assessment's advice).
Participants were also split into prediction groups (which were asked to make quantitative predictions about subjects) and decision groups (which were asked to make binary decisions about subjects).
This design elicits sufficient information to determine whether the risk assessments altered the DMP and is described below in more detail. 

The experimental procedure was the same in both the pretrial and the loans settings. 
After completing a consent page, participants entered a tutorial that described their setting and the predictions or decisions that they would be asked to make.
These descriptions explained the key considerations (including but not limited to risk) that factor into decisions in the relevant setting.
Participants who would be shown the risk assessment's advice were also presented with background information about the risk assessment.
The description of the risk assessment included details about the algorithm's prediction task, training data, and accuracy, and invited participants to use the predictions in whatever manner they desired.  
Participants were unable to proceed beyond the tutorial until they correctly answered several questions demonstrating their comprehension.
We ignored all data from participants who required more than four attempts to correctly answer all of the comprehension questions. 
Participants then completed an intro survey (to provide demographic information and other attributes), a prediction or decision task (described in detail below), and an exit survey (to provide reflections on the task). 

\begin{figure}
	\centering
	\includegraphics[width=\columnwidth]{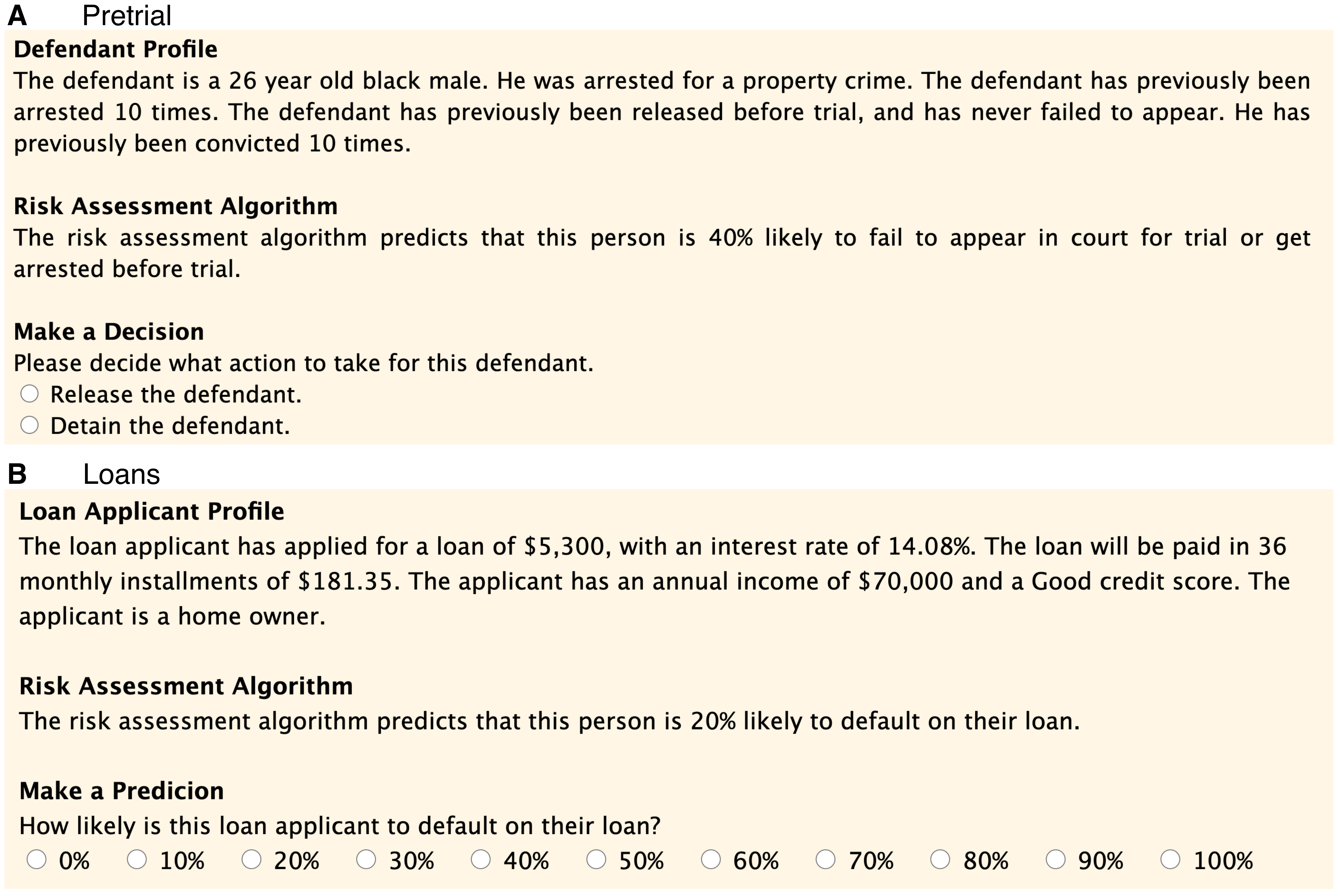}
	\Description{Two examples of prompts presented to participants during the experiment. The prompts include three components. First, we present a several sentence profile describing the defendant or the loan applicant. Second, we show the prediction of the risk assessment for this subject. This second component is presented only for participants in the treatment group. Third, we ask the participant to make a decision or prediction. Decision-making participants make a binary decision. Prediction-making participants are asked to select the option that most closely matches their risk estimate.}
  	\caption{Examples of the prompts presented to participants. 
(A) A profile presented to a decision-making participant in the pretrial setting. 
(B) A profile presented to a prediction-making participant in the loans setting. 
Both of these examples are for participants in the treatment group; participants in the control group saw the same prompt, but without the section about the risk assessment.}
  	\label{fig:prompts}
 \end{figure}

The key component of the experiment was the prediction or decision task (Figure~\ref{fig:prompts}). 
Based on their assigned setting, participants were presented with narrative profiles describing seven features about defendants or applicants.\footnote{These features were the same as those used to develop the risk assessment in each setting, with the addition of race and gender in the pretrial setting.}
These defendants and applicants were drawn randomly from the assigned setting's 300-subject sample. 
Participants were tasked with making either numeric predictions of risk about 40 subjects or binary decisions about 30 subjects.\footnote{Participants making predictions were given a larger number of subjects because they received a bonus payment of up to \$1 in addition to the \$3 base payment received by all participants.}
Prediction-makers were asked to predict risk on a scale from 0\% to 100\%, with options in 10\% increments. 
Decisions in the pretrial setting entailed whether to release or detain criminal defendants before trial.
Decisions in the loans setting entailed whether to approve or reject home improvement loan applications.
This setup matches salient elements of real-world settings such as pretrial adjudication, in which risk assessments are introduced as important decision-making aids \cite{green2020risk, stevenson2018assessing} and in which decisions are often made in just a few minutes \cite{austin2014broward, sji2016report}.

The primary goal of our experiment was to determine the effects of the risk assessments on human decision-making processes.
This requires comparing the decisions of participants with and without a risk assessment while accounting for the risk assessment's effects on predictions. 
We therefore followed a 2x2 experimental setup within each setting, splitting participants according to whether they are presented with the risk assessment and whether they make binary decisions or quantitative risk predictions (Figure~\ref{fig:conditions}). 

Our first experimental condition in each setting was whether or not participants were presented with the predictions of a risk assessment. 
Participants in the control group were shown only the narrative profiles about subjects.
Participants in the treatment group were shown the narrative profiles as well as the risk assessment's predictions about subjects (Figure~\ref{fig:prompts}). 
This first condition allows us to compare the behaviors of participants with and without the risk assessment.

However, directly comparing the decisions of the control and treatment groups cannot determine whether the risk assessment altered the DMP. 
Decisions could differ across the control and treatment groups because the risk assessment influenced the RPP but not the DMP. 
For instance, a risk assessment could increase the likelihood of a defendant being detained before trial by a) making decision-makers more risk-averse or b) causing decision-makers to increase their estimate of the defendant's risk. 
Determining a risk assessment's influence on the DMP therefore requires accounting for the risk assessment's influence on predictions.
This means that we must obtain information regarding participants' perceived risk about subjects in addition to their decisions about subjects.

We could obtain information about the risks perceived by decision-makers in two ways. 
The first approach is to ask each participant to make both predictions and decisions about subjects. 
This approach would provide the most accurate measure of the perceived risk associated with each decision. 
However, because this approach requires directly asking decision-making participants about risk, it would also prime them to consider risk whether or not they are shown the risk assessment.
This priming would undermine the entire study by confounding our ability to detect how presenting a risk assessment influences the consideration of risk in the DMP.
The second approach to measuring perceived risk---which we take in this study---is to have some participants make risk predictions and some participants make decisions.
We use the risk predictions provided by prediction-makers to estimate the risks perceived by decision-makers. 
Although this approach means that we cannot directly measure decision-making participants' perceptions of risk, it provides a reasonable proxy while maintaining the integrity of our research question.

Our second experimental condition, therefore, was whether participants were asked to make predictions or decisions. 
To obtain risk estimates about each subject (both with and without a risk assessment), we asked 75\% of participants to make binary decisions about subjects and 25\% to make numerical predictions of risk about each subject (Figure~\ref{fig:conditions}).\footnote{We placed more participants into the decisions treatment because our analysis required more decisions than predictions to obtain robust results. The disparity in participants making decisions versus predictions is partially offset by the fact that participants making predictions evaluated more subjects (40) than participants making decisions (30).}
We used the risk predictions elicited from the prediction-making participants to estimate the risks perceived by the decision-making participants. 
We estimated the perceived risk associated with a given decision as the average risk prediction made about the subject in question, grouping predictions and decisions based on whether the risk assessment was shown.
For instance, the perceived risk assigned to a decision about a defendant made without the risk assessment was the average of the risk predictions made about that same defendant without the risk assessment. 
By eliciting many predictions about each subject, we obtained reliable measures of the average perceived risk about each subject (both with and without the risk assessment) without inappropriately influencing the behaviors of decision-making participants.

\subsection{Analysis}
To study whether and how a risk assessment alters the decision-making process, we modeled the DMP of participants with and without a risk assessment.
We characterized negative decisions as a function of perceived risk and conducted Bayesian mixed-effects logistic regressions to learn this function.\footnote{We used a Bayesian approach with weak priors to enable analyses based on posteriors. In all cases, the inferences made with Bayesian and non-Bayesian regressions were almost identical.} 
Following the decision-making structure in Figure~\ref{fig:dmp}, we regressed participant decisions on three factors: the perceived risk about the subject in question, whether the risk assessment was shown, and the interaction between these two factors.
Factors such as subject attributes and the risk assessment's prediction are incorporated into this decision function through $perceived.risk$, which is based on these elements.  
We also included three random effects to account for repeated samples in the data.
\begin{equation}
\label{eq:dmp}
\begin{split}
\mathrm{negative.decision} &\sim \mathrm{perceived.risk + show.RA + perceived.risk*show.RA} \\ 
&+ \mathrm{(1|participant) + (1|subject) + (1|progress.index)}
\end{split}
\end{equation}

This regression is structured to infer the DMP that participants followed and to determine whether the risk assessment altered this function, thus distinguishing between Scenario 3 and Scenario 4.
If risk assessments present information that improves the RPP but does not influence the DMP (Scenario 3), we would expect to see that showing the risk assessment does not alter this regression. 
In this case, neither regression factor that includes $show.RA$ would be significant, such that the relationship between decisions and perceived risk is the same whether or not the risk assessment is shown.
However, if risk assessments influence the DMP as hypothesized (Scenario 4), we would expect to see that showing the risk assessment alters this regression, making people more attentive to reducing risk when making decisions. 
This result could emerge through two different mechanisms: 1) the risk assessment makes participants more risk-averse at all levels of risk (in this case, the $show.RA$ coefficient would be positive), or 2) the risk assessment makes participants more sensitive to increases in risk (in this case, the $perceived.risk*show.RA$ coefficient would be positive).

After observing that the risk assessments influenced the DMP (and thus generated Scenario 4 as hypothesized), we estimated the impacts of this influence. 
Our goal was to isolate the effects of the DMP change, controlling for the risk assessments' effects on the RPP.
This analysis entailed comparing outcomes from the observed Scenario 4 behaviors with outcomes from the commonly expected Scenario 3 behaviors. 
Because our control group participants exhibited Scenario 1 and our treatment group participants exhibited Scenario 4, we did not observe Scenario 3 behaviors and could not directly compare Scenario 3 and Scenario 4 outcomes. 
We therefore estimated the differences between Scenario 3 and Scenario 4 outcomes through simulations. 
We began by fitting models for the RPP and DMP in the pretrial and loans settings, both with and without the risk assessment's advice.
We then ran 1,000 trials simulating the outcomes for more than 4,000 defendants and loan applicants in the four scenarios described in Table~\ref{tab:scenarios}.

See Section~\ref{app:analysis} of the Appendix for additional details about our analyses.

\vspace{3mm}
\section{Results}
\subsection{Effects of the Risk Assessments on the Risk-Prediction Process}
\label{sec:results-predictions}
We looked first at how the risk assessments affected predictions of risk.
We evaluated participant ``prediction quality'' using a reverse Brier score bounded between 0 (worst possible performance) and 1 (best possible performance). 
In both settings, presenting the risk assessment reduced estimates of risk, improved prediction accuracy, and aligned the RPP more closely with the risk assessment's calculations. 
These results are consistent with prior work \cite{green2019fat, green2019cscw}.

In the pretrial setting, the risk assessment reduced perceived risk for 54.0\% of defendants.
Overall, defendants received an average reduction in perceived risk of 1.6\% (from 40.6\% to 38.9\%, P=.001, d=0.19). 
While the reduction in perceived risk was significant for white defendants (38.4\% to 35.7\%, P=.003, d=0.30), Black defendants received a smaller and nonsignificant reduction (41.7\% to 40.7\%, P=.085, d=0.12). 
Bayesian linear regression (Equation~\ref{eq:rpp_pretrial}) found that showing the risk assessment altered the risk-prediction process, most notably prompting participants to consider the age of defendants and to reduce the risk associated with violent crime and prior failures to appear (Table~\ref{tab:rpp_shifts}).
Through these changes, presenting the risk assessment increased the average participant prediction quality from 0.72 to 0.75 (P<.001, d=0.11). 

In the loans setting, the risk assessment altered predictions of risk more dramatically. 
The risk assessment reduced the perceived risk for 92.3\% of loan applicants and generated an overall average reduction of 14.2\% for each applicant (from 38.5\% to 24.3\%, P<.001, d=1.54). 
Bayesian linear regression (Equation~\ref{eq:rpp_loans}) found that showing the risk assessment altered the RPP by significantly reducing participants' baseline risk predictions, increasing the salience of annual income and interest rate, and prompting participants to consider the length of loans (Table~\ref{tab:rpp_shifts}).
In turn, showing the risk assessment increased participant prediction quality from 0.75 to 0.83 (P<.001, d=0.31).  

\newpage
\subsection{Effects of the Risk Assessments on the Decision-Making Process}
We next analyzed how the risk assessments affected participant decisions and decision-making processes. 

\subsubsection{Effects on Decisions}
We first compared participant decisions with and without a risk assessment.
Our goal was to investigate whether the shifts in decisions induced by the risk assessments align with the shifts in predictions induced by the risk assessments.
If risk assessments lead to Scenario 3, we would expect to see that shifts in decisions closely track the shifts in predictions described above. 
In particular, reductions in perceived risk due to the risk assessment would be associated with reductions in negative decisions due to the risk assessment, and vice versa. 
Our results do not closely follow this pattern, however, indicating that the risk assessment's effects on decisions cannot be explained by shifts in the RPP alone. 

In the pretrial setting, the risk assessment reduced pretrial detention rates but increased racial disparities. 
The risk assessment reduced each defendant's likelihood of pretrial detention by an average of 2.4\% (from 44.5\% to 42.1\%, P<.001, d=0.21). 
White defendants received a 27\% larger average reduction (38.7\% to 35.9\%, P=.014, d=0.24) than Black defendants (47.7\% to 45.5\%, P=.007, d=0.20).
As a result, the overall racial disparity in pretrial detention increased by 18.8\% from 8.8\% to 10.4\%.
In addition, the risk assessment increased the ``accuracy'' of decisions from 56.7\% to 58.4\% (P=0.009, h=0.03) and reduced the ``false positive rate'' from 26.9\% to 24.4\% (P<.001, h=0.06).\footnote{These measures are described in quotes to reflect that decisions were made with risk as just one of several factors and that it is a simplification to evaluate decisions as if they were predictions of risk.}

In the loans setting, the risk assessment's effects on negative decisions contrasted with the risk assessment's effects on perceived risk. 
Although the risk assessment dramatically reduced risk predictions, the risk assessment did not significantly alter each loan applicant's likelihood of rejection (loan rejection rates went from 22.1\% to 23.1\%, P=.159, d=0.08). 
Furthermore, although the risk assessment significantly increased the accuracy of risk predictions, the risk assessment reduced the ``accuracy'' of decisions from 72.2\% to 70.5\% (P=.002, h=0.04) and increased the ``false positive rate'' from 17.3\% to 18.5\% (P=.015, h=0.03). 
In sum, the risk assessment \emph{notably reduced} perceived risk yet \emph{did not reduce} rejection rates, and similarly \emph{increased} prediction accuracy yet \emph{decreased} decision ``accuracy.'' 

\begin{figure}
	\includegraphics[width=\columnwidth]{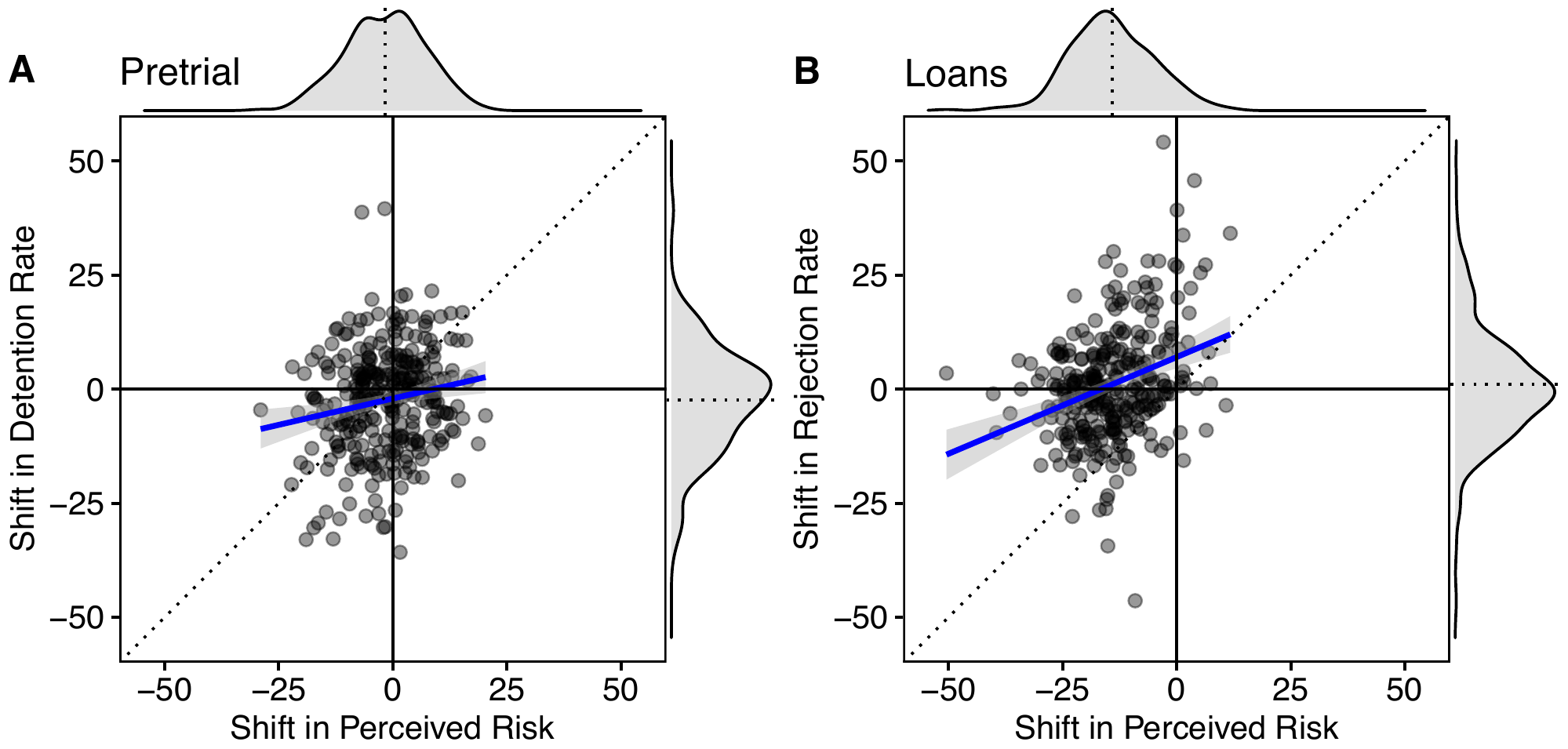}
	\Description{This plot shows how the risk assessment altered the perceived risk and the negative decision rate for each defendant and loan applicant. Both plots demonstrate that although there is a correlation between shifts in perceived risk and shifts in negative decisions, it is relatively weak.}
  	\caption{Shifts in perceived risk and negative decision rates for each subject caused by showing the risk assessment to participants. 
(A) Pretrial setting. (B) Loans setting. 
Each point represents a single defendant or applicant, with marginal density plots along each axis (in which the dotted lines represent the average values). 
Positive values on the x-axis indicate that the risk assessment increased the average risk prediction about a subject. 
Positive values on the y-axis indicate that the risk assessment increased the negative decision rate for a subject. 
Blue lines represent linear regression fits of shifts in negative decisions versus shifts in perceived risk. 
These results indicate that decisions are relatively insensitive to shifts in predictions and that reductions in perceived risk do not necessarily lead to reductions in negative decisions.}
  	\label{fig:shifts}
 \end{figure}

To further investigate the relationship between predictions and decisions, we then compared how the risk assessments altered the predictions and decisions made about each individual subject. 
We found that shifts in perceived risk did not translate to equivalent shifts in negative decisions (Figure~\ref{fig:shifts}).
In both settings, subjects for whom the risk assessment decreased perceived risk did not reliably receive lower negative decision rates due to the risk assessment.
Among the 54.0\% of pretrial defendants for whom the risk assessment reduced perceived risk, only 59.3\% received a reduced likelihood of pretrial detention when the risk assessment was shown.
Among the 92.3\% of loan applicants for whom the risk assessment reduced perceived risk, only 52.0\% received a reduced likelihood of rejection when the risk assessment was shown. 
Overall, shifts in decisions were relatively insensitive to shifts in predictions, with regression coefficients less than 1 in both settings (0.23 in pretrial, P=.003; 0.42 in loans, P<.001). 
For instance, a 10\% reduction in average perceived risk due to the risk assessment was associated with a 4.4\% reduction in the pretrial detention rate and a 2.8\% \emph{increase} in the loan rejection rate. 

These results depart notably from what we would expect to see if the risk assessments induced a shift to Scenario 3. 
These patterns demonstrate that reductions in perceived risk do not lead directly to reductions in pretrial detention or loan application rejections.
Instead, these changes in perceived risk must be mediated through changes to the DMP before yielding decisions. 

\subsubsection{Effects on the Decision-Making Process}
\label{sec:results-decisions}
We next analyzed the risk assessments' effects on the DMP.
Bayesian mixed-effects logistic regressions (Equation~\ref{eq:dmp}) found that the risk assessment altered the decision-making process in both settings, making participants more attentive to risk when making decisions.
These results demonstrate that the risk assessments prompted the hypothesized shift to Scenario 4 rather than Scenario 3. 

\begin{figure}
	\centering
	\includegraphics[width=.95\columnwidth]{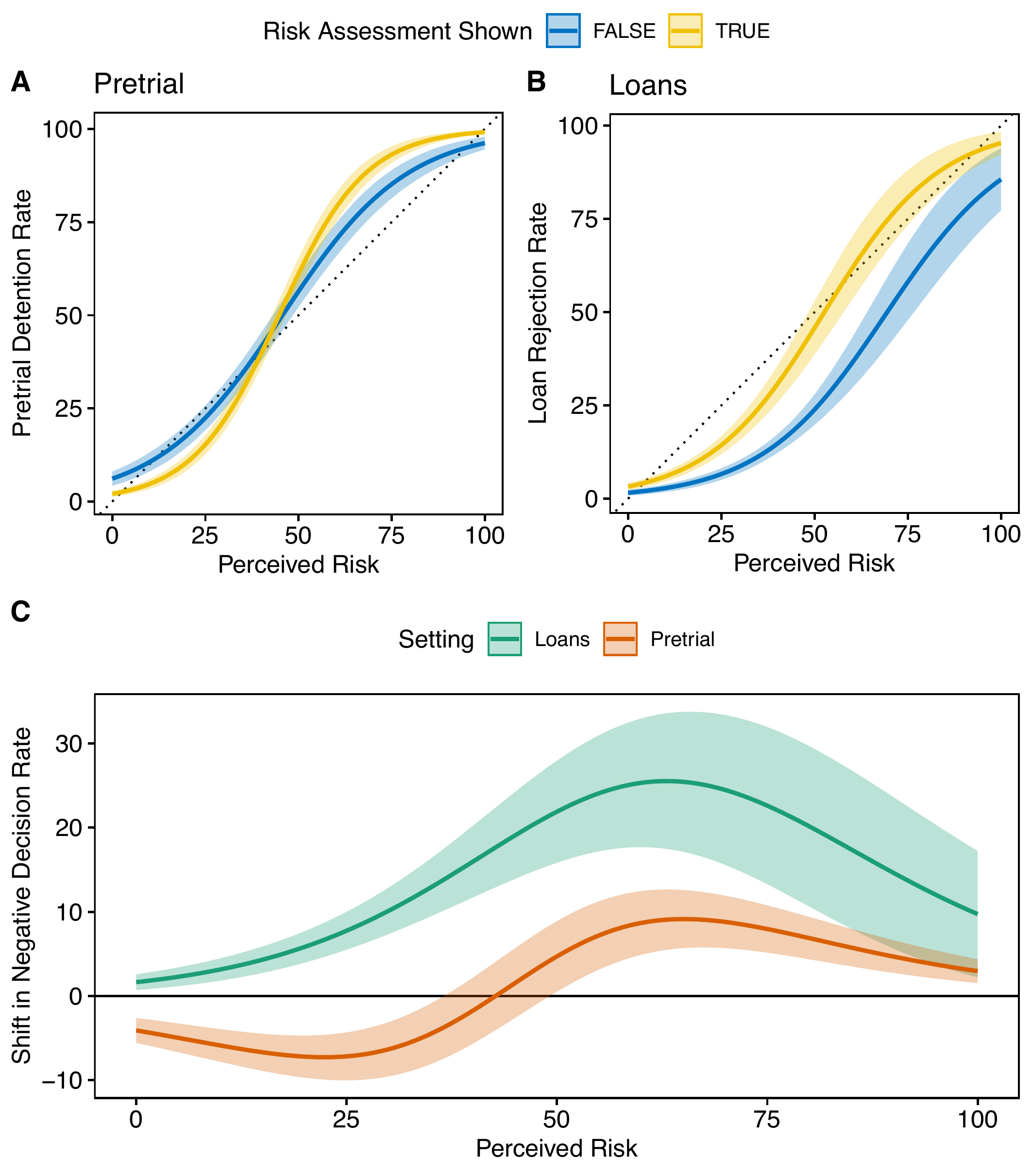}
	\Description{This plot summarizes how showing the risk assessment alters decision-making. Subplots A and B have lines showing the human decision-making function with and without the risk assessment's advice. Subplot C shows the difference in negative decision rates between the decision-making functions with and without the risk assessment, as a function of perceived risk. One line shows the effects in the loans setting and another line shows the effects in the pretrial setting.}
  	\caption{Changes in the decision-making processes caused by showing the risk assessments to participants. 
(A) Decision functions indicating the likelihood of detaining a pretrial defendant based on the perceived risk of that defendant (see Table~\ref{tab:dmp_shifts} for model coefficients). 
The risk assessment made participants more sensitive to increases in perceived risk, reducing detention at low risk and increasing detention at high risk. 
(B) Decision functions indicating the likelihood of rejecting a loan application based on the perceived risk of that applicant (see Table~\ref{tab:dmp_shifts} for model coefficients). 
The risk assessment caused rejection rates to increase at all levels of perceived risk. 
(C) Shift in negative decision (i.e., pretrial detention or loan rejection) probability due to the shift in the DMP caused by showing the risk assessment. 
Given a perceived risk of 50\%, for instance, the DMP shift increased the likelihood of pretrial detention by 4.7\% and the likelihood of loan rejection by 21.9\%. 
Bands indicate 95\% confidence intervals in all panels.
The values behind this figure are summarized in Table~\ref{tab:decision_probs}.}
  	\label{fig:dmp_shifts}
 \end{figure}

In the pretrial setting, the risk assessment made participants more sensitive to increases in risk (Figure~\ref{fig:dmp_shifts}). 
Presenting the risk assessment increased the odds ratio associated with a 10\% increase in perceived risk from 1.82 to 2.39 (Table~\ref{tab:dmp_shifts}).
These results mean that the risk assessment made perceived risk a stronger determinant of whether defendants were released or detained: the risk assessment reduced pretrial detention rates for defendants with low perceived risk and increased pretrial detention rates for defendants with high perceived risk. 
For example, the risk assessment reduces the detention likelihood by 6.3\% for a defendant with a perceived risk of 30\% but increases the detention likelihood by 8.7\% for a defendant with a perceived risk of 60\% (Table~\ref{tab:decision_probs}).

In the loans setting, the risk assessment made participants more risk-averse at all levels of risk (Figure~\ref{fig:dmp_shifts}). 
Presenting the risk assessment increased the odds of rejecting loan applications by a factor of 2.09 (Table~\ref{tab:dmp_shifts}). 
For all levels of perceived risk up to 46.0\% (covering 97.3\% of risk estimates with the risk assessment), participants were more than twice as likely to reject loan applications if they were shown the risk assessment (Table~\ref{tab:decision_probs}). 
For instance, an applicant with a perceived risk of 30\% would have an 8.7\% likelihood of being rejected by a participant not shown the risk assessment and an 18.8\% likelihood of being rejected by a participant shown the risk assessment. 

\begin{table}[]
\caption{Bayesian mixed-effects logistic regression results estimating the likelihood of a negative decision about defendants and loan applicants as a function of perceived risk, following Equation~\ref{eq:dmp}.
The first column presents the coefficient of each factor; the second column presents the coefficient of the interaction between that factor and the risk assessment being shown. 
The second column thus describes how showing the risk assessment altered each factor.
Parenthetical terms represent standard errors and terms in brackets represent odds ratios. 
The intercept represents modeled participant responses at a perceived risk of 0\%, with perceived risk measured in units of 10\%. 
In the pretrial setting, presenting the risk assessment reduced the likelihood of detention for 0\% risk but increased participants' sensitivity to increases in risk.
In the loans setting, presenting the risk assessment increased the odds of rejecting loan applications by a factor of 2.09. 
These patterns are plotted in Figure~\ref{fig:dmp_shifts}.
. P<0.1; * P<0.05; ** P<0.01; *** P<0.001
}
\begin{tabular}{@{}lll@{}}
\toprule
 & Not Shown RA & Shown RA (interaction) \\ \midrule
\textbf{Pretrial} &  &  \\
Intercept & –2.74 (0.17) *** & –1.14 (0.14) [0.32] *** \\
Perceived Risk & 0.60 (0.04) [1.82] *** & +0.27 (0.03) [1.31] *** \\
 &  &  \\
\textbf{Loans} &  &  \\
Intercept & –4.15 (0.24) *** & +0.74 (0.22) {[}2.09{]} *** \\
Perceived Risk & 0.60 (0.05) [1.82] *** & +0.05 (0.05) [1.05] \\ \bottomrule
\end{tabular}
\label{tab:dmp_shifts}
\end{table}

\newpage
When asked to reflect on their behavior after making decisions, participants did not seem to recognize that the risk assessment had altered how they consider risk when making decisions. 
Despite becoming more attentive to risk when making decisions, participants presented with a risk assessment expressed less support for basing decisions on risk (Pretrial: P=.003, d=0.21; Loans: P=.001, d=0.23). 
Furthermore, the risk assessments did not alter participant reports regarding the priority that decision-makers should assign to key considerations such as risk (Table~\ref{tab:priorities}).

\subsection{Impacts of the Shifts in the Decision-Making Process}
\label{sec:results-simulations}
We used simulations to estimate the impacts of each risk assessment's influence on the DMP.
Our goal was to isolate the effects of the DMP shifts by controlling for the concurrent RPP shifts.
We accomplished this through simulations that enabled us to compare the observed Scenario 4 outcomes with the commonly expected Scenario 3 outcomes.

\begin{figure}
	\centering
	\includegraphics[width=\columnwidth]{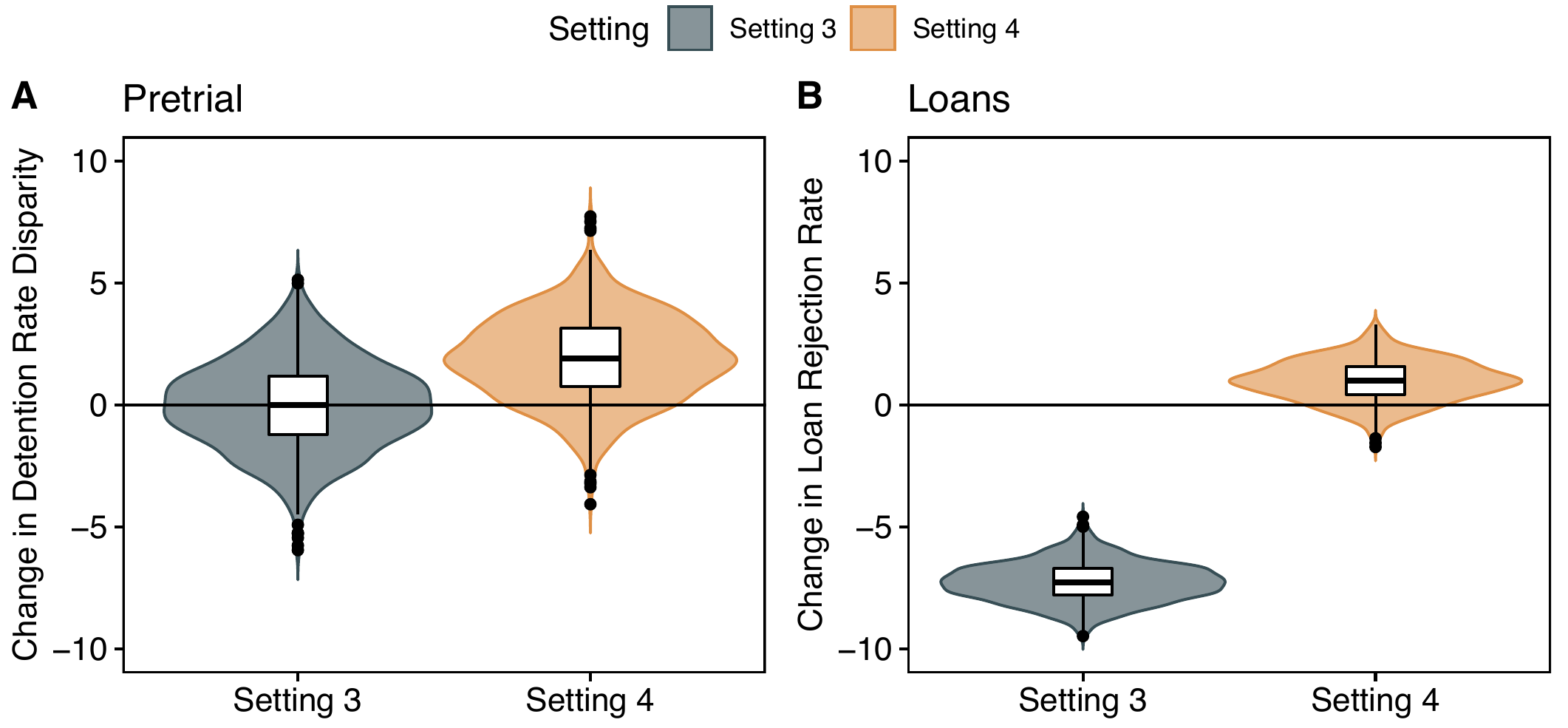}
	\Description{This figure shows the effects of the DMP shift, comparing Scenarios 3 and 4. Subplot A is for the pretrial setting, showing that Setting 4 increased the racial disparity in detention rate relative to Setting 3. Subplot B is for the loans setting, showing that Setting 4 significantly increased the loan rejection rate relative to Setting 3.}
  	\caption{Simulated outcomes in Scenarios 3 and 4 compared to Scenario 1. 
(A) Change in the Black-white detention rate disparity in the pretrial setting. 
Scenario 3 did not significantly alter the average racial disparity while Scenario 4 increased the average racial disparity by 1.9\%. 
The shift in the DMP is therefore responsible for a 1.9\% increase in racial disparities.
(B) Change in the rejection rate in the loans setting. 
Scenario 3 reduced the average rejection rate by 7.3\% while Scenario 4 increased the average rejection rate by 1.0\%. 
The shift in the DMP is therefore responsible for an 8.3\% increase in loan rejections.}
  	\label{fig:simulations}
 \end{figure}

In the pretrial setting, the risk assessment's influence on the DMP reduced the average detention rate but exacerbated racial disparities (Figure~\ref{fig:simulations}).
Had the risk assessment affected only the RPP (i.e., created a shift from Scenario 1 to Scenario 3), none of the ``accuracy,'' ``false positive rate,'' nor detention rates for either race would have changed. 
The shift in the DMP (i.e., from Scenario 3 to Scenario 4) increased decision ``accuracy'' from 57.7\% to 60.4\% (P<.001, d=4.43), decreased the ``false positive rate'' from 27.4\% to 24.2\% (P<.001, d=6.20), and reduced detention by 4.9\% for white defendants and by 3.0\% for Black defendants (P<.001, d=1.52). 
Thus, although the DMP shift improved some outcomes, it also increased the racial disparity by 1.9\% and by a factor of 1.34 from 5.6\% in Scenario 3 to 7.5\% in Scenario 4 (P<.001, d=1.06; Figure~\ref{fig:simulations}).

In the loans setting, the change in the DMP caused by the risk assessment generated a notable decrease in ``accuracy'' and increase in rejections (Figure~\ref{fig:simulations}). 
Had the risk assessment affected only the RPP and thus prompted a shift from Scenario 1 to Scenario 3, the decision ``accuracy'' would have increased from 70.8\% to 75.6\% (P<.001, d=8.82), the ``false positive rate'' would have decreased from 17.5\% to 11.5\% (P<.001, d=12.31), and the rejection rate would have dropped from 22.2\% to 14.9\% (P<.001, d=13.09). 
The shift in the DMP negated these potential benefits, however, as the risk assessment made participants more risk-averse.
Moving from Scenario 3 to Scenario 4 decreased the decision ``accuracy'' from 75.6\% to 70.7\% (P<.001, d=8.83), increased the ``false positive rate'' from 11.5\% to 18.1\% (P<.001, d=13.26), and increased the rejection rate from 14.9\% to 23.2\% (P<.001, d=14.88). 
Overall, instead of simply improving risk predictions and thereby generating a 7.3\% increase in loans granted, the risk assessment also increased risk-aversion and thereby actually \emph{reduced} the loans granted by 1.0\% (Figure~\ref{fig:simulations}).
The shift in the DMP is therefore responsible for an 8.3\% increase in loan rejections. 

\section{Discussion}
This paper provides the first direct evidence that risk assessments can systematically alter how people balance risk with other factors when making policy-relevant decisions. 
Even though our risk assessments improved the accuracy of human predictions, they also induced shifts in decision-making processes that counteracted the potential benefits of these improved predictions. 
Presenting a risk assessment increased participant sensitivity to risk in pretrial detention decisions (thus exacerbating racial disparities) and increased participant risk-aversion in government loan decisions (thus reducing the loans granted). 
These shifts mean that even when the risk assessments reduced participant predictions of risk about subjects, participants did not accordingly reduce the rate of negative decisions about those subjects.
Alternative explanations, such as the risk assessments simply making participants more confident in their risk estimates, can be ruled out by our data (see Section~\ref{app:alternatives} in the Appendix).

\subsection{Policy Implications}
Our results challenge the assumption that improving human predictions with risk assessments will necessarily improve human decision-making---an assumption that has been central to the adoption of algorithmic decision-making aids by governments.
These findings demonstrate the potential limits and harms of efforts to improve public policy by incorporating predictive algorithms into multifaceted policy decisions. 
If the observed changes were to occur in real-world settings, they would be notable for three primary reasons. 

First, our findings indicate that government algorithms could generate unexpected shifts in public policy and jurisprudence. 
Although improving the accuracy of risk predictions is consistent with policies that include risk as a consideration, a systematic increase in the salience of risk amounts to a shift in the normative balancing act that comprises public policy in domains such as pretrial adjudication \cite{koepke2018danger, mahoney2001pretrial}.
Such a shift reduces the range of factors that decision-makers consider, diminishing the implementation of public policy \cite{zacka2017state}.
In pretrial settings, increasing the weight that judges place on risk would generate undue social harms \cite{yang2017optimal} and enhance the constitutionally contested policy of preventative detention (detaining defendants until trial due to their likelihood to commit future crimes) \cite{koepke2018danger, green2020risk}.
In loans settings, greater risk-aversion would reduce government aid and would counteract the goal of promoting equity through giving loans to low-income (and hence high-risk) applicants.

Second, because risk is intertwined with legacies of racial discrimination in the criminal justice and financial systems, more heavily basing decisions on risk would likely exacerbate racial disparities in incarceration and government aid. 
Due to past and present oppression in the United States, Blacks have disproportionately higher risk levels than whites for being arrested and defaulting on loans, making them particularly vulnerable to increased attention to risk \cite{green2020risk, kiviat2019credit}. 
Indeed, we found that the DMP shifts caused by the risk assessments increased the racial disparity in pretrial decisions and reduced government aid in loans decisions. 

Third, because these two effects would arise as an unexpected byproduct of integrating an algorithm into decision-making, they would occur without deliberation or oversight.
These shifts in policy and jurisprudence (and the resulting racial disparities) would be the consequence of an algorithm's unintended influence on human decision-making rather than a democratic policymaking process.
Because these effects are unexpected, they would likely evade scrutiny, at least until their effects manifest in practice with sufficient evidence. 
Such changes would likely be further obscured by decision-makers not recognizing that the risk assessment had influenced their behavior, as observed both here and in prior work \cite{green2019fat, green2019cscw}.
These effects add another dimension to the unexpected and unaccountable policy distortions that emerge when laws are translated into code \cite{citron2008process}.

Together, these implications highlight harms that can arise when algorithms are incorporated into multifaceted policy decisions. 
If evaluations of algorithmic decision-making aids do not account for human-algorithm interactions and the many normative considerations relevant to policy decisions, they are likely to overestimate the benefits and underestimate the harms of incorporating algorithms into government decision-making \cite{green2020realism, stevenson2019hands}.

\subsection{Future Work}
There is an urgent need to uncover potential issues in human-algorithm collaborations \emph{before} algorithms shape life-changing decisions. 
Risk assessments are increasingly being integrated into high-stakes decisions, yet consistently produce unexpected and unjust impacts in practice \cite{stevenson2018assessing, stevenson2019hands, albright2019judge, brayne2020technologies}.
Achieving a more responsible approach to algorithm-in-the-loop decision-making requires several areas of future work.

\subsubsection{Open Research Questions}
It is necessary to develop a deeper scientific understanding of how risk assessments and other algorithms influence human decision-making. 
Although we demonstrated that presenting risk assessments can alter human decision-making processes in harmful ways, many open questions remain.

One open question is how the effects of risk assessments vary across contexts. 
Notably, our risk assessments exerted different effects across the two settings studied, making participants more sensitive to increases in perceived risk in the pretrial setting and more risk-averse in the loans settings.
We do not know what caused the observed differences across the two settings.
One hypothesis is that the effects of a risk assessment depend on people's pre-existing notions of risk in that context.
For instance, people may be strongly predisposed to consider risk in pretrial decisions, such that the risk assessment merely amplified this behavior, but not in government loans decisions, such that the risk assessment prompted heightened concern about mitigating risk.

An important role for future inquiry will be to study how algorithms alter decision-making processes in different settings and with different decision-makers.
Algorithms are being deployed in many social contexts beyond government, such as schools \cite{holstein2019classrooms}, hospitals \cite{kiani2020cancer}, and newsrooms \cite{christin2020metrics}.
Although these settings involve some straightforward prediction problems, in many cases people must integrate predictions with other considerations to make decisions. 
Determining the proper roles for algorithms in these and other settings thus requires a deeper understanding of how algorithmic predictions influence human decision-making across contexts. 

A second open question is whether any mechanisms could mitigate the risk assessments' effects on decision-making processes, such that these algorithms do in fact lead to the widely expected Scenario 3 outcomes.
It is possible that other approaches to presenting algorithms and structuring decision-making could improve how people incorporate algorithmic advice into their decisions. 
In the context of algorithm-aided human predictions, for instance, asking people to make preliminary predictions before being shown a risk assessment's predictions modestly improved accuracy and fairness, whereas providing feedback and explanations did not improve performance \cite{green2019cscw}.

\subsubsection{Developing a Proactive Pipeline of Evaluations}
It is also necessary to develop a testing pipeline that evaluates human interactions with algorithmic decision-making aids before these tools are implemented in practice.  
Decisions to adopt algorithms should require a baseline of evidence suggesting that they are actually likely to improve decision-making.
Our results show that a central assumption motivating risk assessments in public policy---that improving human predictions will improve human decisions---can be violated with laypeople.
This finding suggests the need to investigate whether this assumption holds in practice.
Furthermore, many regulations across the world point to human oversight as providing protections against algorithms, yet these protections rarely function as desired \cite{green2021slate}. 
Rather than relying on untested assumptions, efforts to integrate algorithms into public policy should be grounded in proactive evaluations of proposed human-algorithm collaborations.

Attaining more thorough knowledge about the effects of algorithmic decision-making aids will require a pipeline of evaluations that combines several modes of analysis: experimental studies with laypeople in lab settings, experimental studies with domain experts in lab settings, and ethnographic and empirical studies of expert interactions with algorithms in practice.
Each of these modes has particular strengths and weaknesses.
Collectively, they can provide robust, proactive knowledge about how algorithms affect human decision-making and how to improve human-algorithm collaborations. 

Developing this pipeline requires further exploring how public servants collaborate with algorithms in practice and how lab experiments can inform the implementation of algorithmic decision-making aids. 
The primary limitation of this paper is that our findings are based on the behaviors of Mechanical Turk workers in a lab experiment rather than judges or loan agents operating in real-world contexts.
Our results do not directly reflect how algorithms affect the behaviors of experts making real decisions. 
There are likely to be significant differences between how laypeople and trained experts make decisions with algorithms, particularly related to perceptions of professional identity and autonomy \cite{brayne2020technologies}. 

Despite these differences, experiments with laypeople can shed light on some behaviors of experts in practice. 
Research suggests that both judges \cite{guthrie2001inside, rachlinski2008judges, rachlinski2018framing} and financial professionals \cite{cohn2015evidence, gilad2008priming, haigh2005professional} are susceptible to priming and framing effects (alongside other cognitive biases) in much the same manner as laypeople. 
Prior studies of how laypeople interact with risk assessments \cite{green2019fat, green2019cscw} have demonstrated racially biased behaviors similar to those observed among judges using risk assessments in practice \cite{albright2019judge, cowgill2018impact}.
Furthermore, the results of this study align with prior experiments suggesting that risk assessments cause law students and judges to place a greater priority on reducing crime risk \cite{skeem2019impact, starr2014sentencing} and that pretrial risk assessments have increased racial disparities in practice \cite{albright2019judge, stevenson2018assessing}. 

Lab studies with laypeople therefore present a valuable approach for attaining preliminary insights about human-algorithm collaborations.
Initial trials with laypeople can provide a foundation of knowledge about how algorithms influence decision-makers and whether there are mechanisms that can improve these collaborations.
Compared to studies with experts, experimental studies with laypeople have several advantages. 
Most importantly, such experiments allow us to learn about human-algorithm collaborations before implementing an algorithm into real-world contexts. 
Furthermore, compared to lab and in situ evaluations with practitioners, experiments with laypeople can be conducted more quickly, with more participants, and with more precisely controlled experimental procedures. 
Insights from experiments with laypeople can inform the hypotheses and methods for studies with practitioners, which provide more precise knowledge about human-algorithm collaborations in a particular context but are more intensive to run.\footnote{Studies with practitioners can also evaluate the validity of experiments with laypeople and determine what kinds of knowledge such experiments can and cannot reliably provide.}

A proactive pipeline of evaluations along these lines should become a central component of proposals and policies for how governments use algorithmic decision-making aids. 
If algorithms such as risk assessments are to be implemented in a given policy context, there must first be rigorous evidence regarding what impacts they are likely to generate and democratic deliberation supporting those impacts.

\begin{acks}
We thank the area chairs and reviewers for thoughtful feedback regarding how to improve the manuscript. 
We also thank Alan Altshuler, Evan Green, Ben Lempert, and Salom\'{e} Viljoen for their helpful comments on earlier drafts of this manuscript and Steve Worthington for consultation on statistical methodology. 
This material is based upon work supported by the National Science Foundation Graduate Research Fellowship Program under Grant No. DGE1745303. 
This work was also supported by the Michigan Society of Fellows.
\end{acks}

\bibliographystyle{ACM-Reference-Format}
\bibliography{references.bib}

\par\bigskip\noindent\small\normalfont
Received January 2021; revised April 2021; accepted July 2021
\par

\newpage
\appendix

\renewcommand{\thefigure}{A.\arabic{figure}}
\setcounter{figure}{0}

\renewcommand{\thetable}{A.\arabic{table}}
\setcounter{table}{0}

\renewcommand{\theequation}{A.\arabic{equation}}
\setcounter{equation}{0}

\section{Data and Risk Assessments}
\label{app:RAs}
\subsection{Pretrial Detention}
To create our pretrial risk assessment, we used the dataset ``State Court Processing Statistics, 1990-2009: Felony Defendants in Large Urban Counties,'' which was collected by the U.S. Department of Justice \cite{doj2009icpsr}. 
The dataset contains court processing information about 151,461 felony cases filed in May in even years from 1990-2006 and in 2009 in 40 of the 75 most populous counties in the United States. 
The data contains information about each case that includes the arrest charges, the defendant's demographic characteristics and criminal history, and the outcomes of the case related to pretrial release (whether the defendant was released before trial and, if so, whether they were rearrested before trial or failed to appear in court for trial).

We first cleaned the dataset. 
We removed incomplete entries and restricted our analysis to defendants who were at least 18 years old and whose race was recorded as either Black or white. 
In order to have ground-truth data about whether a defendant was rearrested before trial or failed to appear for trial, we also restricted our analysis to defendants who were released before trial.

This yielded a dataset of 47,141 defendants (Table~\ref{tab:defendants}). 
The defendants were primarily male (76.7\%) and Black (55.7\%), with an average age of 30.8 years. 
Among these defendants, 15.0\% were rearrested before trial, 20.3\% failed to appear for trial, and 29.8\% exhibited at least one of these outcomes (which we defined as ``violating'' the terms of pretrial release).

We used this data to train a risk assessment (i.e., a machine learning classifier) that predicts whether each defendant will violate pretrial release. 
We trained the model using gradient boosted decision trees \cite{friedman2001greedy} with the \verb|xgboost| implementation in \verb|R| \cite{chen2020xgboost}. 
The classifier incorporated five features about each defendant: age, offense type, number of prior arrests, whether that person has any prior failures to appear, and number of prior convictions. 
Despite knowing the race and gender of defendants, we excluded these attributes from the model to match common practice among risk assessment developers \cite{arnold2019faqs}.

We performed model selection and evaluated the model using ten-fold cross-validation. 
We first set aside a random sample of 10\% of the data as a held-out validation set.
We then took the remaining 90\% of the data as the training data. 
We split this training data into ten folds, using cross-validation to find hyperparameters for the boosted trees model. 
Cross-validation on the final model yielded an average test AUC of 0.66 (sd=0.009). 
We then trained the model on the complete training data and applied it to the held-out validation set, yielding an AUC of 0.67. 
This indicates comparable accuracy to COMPAS \cite{larson2016compas}, the Public Safety Assessment \cite{demichele2018psa}, and other risk assessments used in practice \cite{desmarais2013risk}.

We selected a sample of 300 defendants from the validation set whose profiles would be shown to participants during the Mechanical Turk experiment. 
To protect defendant privacy, this sample could include only defendants whose seven displayed attributes were shared with at least two other defendants in the complete dataset. 
This restriction meant that we could not select a uniform random sample of 300 defendants from the validation set.
However, we found in practice that sampling from the validation set with weights based on each defendant's risk score yielded a sample population that resembles the complete set of released defendants across most dimensions (Table~\ref{tab:defendants}).

\subsection{Home Improvement Loans}
We used a dataset of loans from the peer-to-peer lending company Lending Club to create our loans risk assessment. 
The data contains records about all 2,004,091 loans that Lending Club issued between 2007 and 2018. 
Each record includes information such as the purpose of the loan; the loan applicant's job, annual income, and approximate credit score; the loan amount and interest rate; and whether the borrower paid off the loan. 
The data includes the first three digits of each borrower's zip code but does not include further demographic information (such as the age, race, or gender of applicants).

We cleaned the dataset to remove incomplete entries and classified credit scores into one of five categories (Poor, Fair, Good, Very Good, and Exceptional), as defined by FICO \cite{fico2016scores}. 
We restricted our analysis to loans issued for home improvements, which represents 6.7\% of the total issued loans.
Home improvement loans represent the third most common purpose for loans in the dataset, following debt consolidation and paying off credit cards. 
We also limited the data to loans that have been either fully paid or defaulted on.\footnote{Although the data represents unpaid loans as being ``charged off,'' which is more extreme than defaulting on a loan, we refer to charged off loans as being defaulted on because the latter is the more commonly used and understood term.}

This yielded a dataset of 45,218 home improvement loans (Table~\ref{tab:loans}). 
The average loan was for \$14,556.38.
The average applicant had an income of \$95,262.88 and a credit score of 707.5 (categorized by FICO as ``Good''). 
More than 80\% of these loans were fully paid off.

We used this data to train a risk assessment that predicts whether each loan will be defaulted on. 
We trained the classifier using gradient boosted decision trees \cite{friedman2001greedy} with the \verb|xgboost| implementation in \verb|R| \cite{chen2020xgboost}. 
Our model considered seven factors about each loan: three factors about each applicant (annual income, credit score category, and whether they own their home) and four factors about each loan (total value, interest rate, monthly installment, and whether its repayment term is 36 or 60 months).

We evaluated the model using ten-fold cross-validation, following the procedure described above for the pretrial risk assessment. 
Cross-validation on the final model yielded an average test AUC of 0.70 (sd=0.01). 
Training the classifier on the complete training data (90\% of the samples) and applying it to the held-out validation set (the remaining 10\% of the data) yielded an AUC of 0.69. 
This performance is similar to that of other loan default risk assessments \cite{ustun2019recourse}.

We selected a sample of 300 loan applicants from the validation set whose profiles would be shown to participants during the Mechanical Turk experiment (Table~\ref{tab:loans}).
These applicants were selected through a uniform random sample from the complete validation set.

\section{COVID-19 Reliability Analysis}
\label{app:covid}
As we prepared to run our experiment in May 2020, we wanted to ensure that our results would not be the product of aberrant behavior prompted by the COVID-19 pandemic.
Before running the full experiment, therefore, we conducted a retest of a trial experiment that we had conducted in December 2019. 

The December 2019 trial closely resembled the experiment described in the main text. 
We recruited 240 participants from Mechanical Turk to evaluate a sample of 100 defendants. 
For the May 2020 trial, we recruited 250 participants to evaluate the same set of 100 defendants. 
We compared the results of these two trials to determine whether COVID-19 altered the population of Mechanical Turk workers or human interactions with risk assessments.
We focused on three results central to our study: the demographics of participants, how participants made risk predictions, and how participants made decisions about whether to release or detain defendants.
For all three results, we did not observe any notable differences across the two trials, suggesting that COVID-19 did not have notable impacts on our results.

\subsection{Participant Demographics}
The demographics of our study participants were similar across the two trials. 
In both cases, participants were predominantly white (80.5\% in 12/2019 vs. 73.4\% in 05/2020), male (58.6\% vs. 58.0\%), and college-educated (73.5\% vs. 70.2\%). 
A logistic regression predicting which trial each participant was part of (based on all of the demographic attributes reported during the intro survey) yielded no terms that were statistically significant.

\subsection{Predictions}
We observed a high degree of consistency between the predictions made across the two trials. 
The correlation between the average prediction made about each of the 100 defendants was r(198)=+.94, P<.001. 
A two-sided t-test yielded no statistically significant difference between participants' prediction quality across the two trials (0.751 vs. 0.753, P=.820).

We also estimated the function used by participants to predict the risk of each defendant. 
We used a mixed-effects linear regression model to measure the average risk prediction about each defendant, grouped by whether the risk assessment was shown and whether the prediction was made in the first or second trial (we refer to this variable as ``trial number''). 
The model included fixed effects for whether the risk assessment was shown, whether the predictions were made in the first or second trial, the attributes of defendants, and the interactions between these three sets of factors (up to three-way). 
We also included random effects for participant and defendant identities to account for repeated samples. 
Our goal was to evaluate whether trial number influenced how participants made predictions. 
We observed minimal differences in the prediction function used across the two trials. 
The trial number and the interaction between trial number and whether the risk assessment was presented were not statistically significant. 
Only two of the interactions that included trial number were statistically significant: participants were slightly less responsive to prior failures to appear (P=.025) and prior convictions (P=.039) in the second trial.

\subsection{Decisions}
Finally, we  observed a high degree of consistency between the decisions made across the two trials.
The correlation between the average detention rate for each of the 100 defendants was r(198)=+.97, P<.001. 

We also estimated the function used by participants to decide whether to release or detain each defendant. 
We used a mixed-effects logistic regression model on all 8,070 decisions made across the two trials. 
The model included fixed effects for whether the risk assessment was shown, the trial number, and the perceived risk about each defendant, with up to three-way interactions between these factors. 
We included random effects for participants, defendants, and status in the experiment to account for repeated measurements. 
None of the coefficients that included trial number were statistically significant, indicating that the decision-making function did not notably differ across the December 2019 or the May 2020 trials.

\subsection{Summary}
In sum, we found high levels of test-retest reliability.
The results found in May 2020 (in the early stages of the COVID-19 pandemic) closely resemble the results found in December 2019.
This suggests that the results presented in this paper were not notably influenced by aberrant behaviors that arose in response to COVD-19. 
More broadly, this also indicates the reliability of our results as being reproducible upon repeated experimentation.

\section{Analysis}
\label{app:analysis}
\subsection{Predictions}
This section provides further detail on the results provided in Section~\ref{sec:results-predictions}.
We estimated the risk-prediction process of participants using Bayesian linear regression. 
We used a Bayesian approach for consistency with the next section, where Bayesian regression enabled analysis based on posteriors.
For all results throughout the paper, the inferences made from Bayesian and non-Bayesian regressions were almost identical.
We implemented models with the \verb|brms| package in \verb|R| \cite{burkner2018brms}, which provides a high-level interface to Markov Chain Monte Carlo (MCMC) sampling for Bayesian inference using \verb|Stan| \cite{carpenter2017stan}. 

In both settings, we regressed the average prediction about each subject (both with and without the risk assessment) on the subject attributes presented to participants and a binary variable ($show.RA$) reflecting whether the risk assessment was shown. 
We included interactions between $show.RA$ and each subject attribute.
To account for repeated samples of subjects, the model also included random effects for subject identity. 
Equation~\ref{eq:rpp_pretrial} is the regression in the pretrial setting and Equation~\ref{eq:rpp_loans} is the regression in the loans setting.

\begin{equation}
\label{eq:rpp_pretrial}
\begin{split}
\mathrm{perceived.risk} &\sim \mathrm{race + gender + age + offense.type + number.prior.arrests} \\
&+ \mathrm{prior.failure.to.appear + show.RA + race*show.RA}  \\
&+ \mathrm{gender*show.RA + age*show.RA + offense.type*show.RA} \\
&+ \mathrm{number.prior.arrests*show.RA + number.prior.convictions*show.RA}  \\
&+ \mathrm{prior.failure.to.appear*show.RA + (1|subject)}
\end{split}
\end{equation}

\begin{equation}
\label{eq:rpp_loans}
\begin{split}
\mathrm{perceived.risk} &\sim \mathrm{income} + \mathrm{fico.category} + \mathrm{own.home} + \mathrm{monthly.installment} \\
&+ \mathrm{interest.rate} + \mathrm{loan.amount} + \mathrm{loan.term} + \mathrm{show.RA} \\
&+ \mathrm{income*show.RA} + \mathrm{fico.category*show.RA} + \mathrm{own.home*show.RA} \\
&+ \mathrm{monthly.installment*show.RA} + \mathrm{interest.rate*show.RA} \\
&+ \mathrm{loan.amount*show.RA} + \mathrm{loan.term*show.RA} + (1|\mathrm{subject})
\end{split}
\end{equation}

We initialized models with uninformative priors and implemented sampling using four chains with 1,000 iterations, following 1,000 burn-in iterations on each chain. 
All coefficients in both models returned $\hat R=1.00$, indicating that the chains were well-mixed and converged to a common distribution. 
We estimated statistical significance from the samples by using the probability of direction measure and obtaining the equivalent frequentist p-value \cite{makowski2019indices, makowski2019bayestestr}. 
The results are summarized in Table~\ref{tab:rpp_shifts}.

\subsection{Decisions}
This section provides further detail on the results provided in Section~\ref{sec:results-decisions}.
We estimated the decision-making process using Bayesian mixed-effects logistic regression, implemented in \verb|brms| \cite{burkner2018brms}. 
In both settings, we regressed each decision on the perceived risk about the subject in question, whether the risk assessment was shown, and the interaction between these two factors (Equation~\ref{eq:dmp}).\footnote{The $perceived.risk$ measurements are based on an average of 18.13±4.00 participant predictions about each subject in each treatment (RA or no RA), with an average standard deviation in risk predictions of 21.85±6.64 and an average standard error of 5.21±1.70. These values are almost identical across the two settings.}
To account for repeated samples, the model also included random effects for the participant identity, the subject identity, and the progress index (1--30) marking the participant's progress in the experiment. 

We initialized models with uninformative priors and implemented sampling using four chains with 1,000 iterations, following 1,000 burn-in iterations on each chain. 
In both settings, all fixed effect coefficients returned $\hat R=1.00$ and all random effect coefficients returned $\hat R \leq 1.01$, indicating that the chains were well-mixed and converged to a common distribution. 
We estimated statistical significance from the samples by using the probability of direction measure and obtaining the equivalent frequentist p-value \cite{makowski2019indices, makowski2019bayestestr}. 
The results are summarized in Table~\ref{tab:dmp_shifts}. 
The standard deviations for the random effects in the pretrial setting are 1.03 for worker, 0.90 for subject, and 0.07 for experiment progress index. 
The standard deviations for the random effects in the loans setting are 1.19 for worker, 0.90 for subject, and 0.29 for experiment progress index. 

We then studied the characteristics of these fitted decision-making process functions (Figure~\ref{fig:dmp_shifts} and Table~\ref{tab:decision_probs}).  
We did this using all 4,000 posterior samples of the fixed effect coefficients from the fitted model.
First, we used these samples to calculate the fitted negative decision rate at each level of risk from 0\% to 100\% (in intervals of 0.1\%), both with and without the risk assessment. 
Second, we used these posterior estimates to calculate the shifts in negative decision rates caused by the risk assessment at each level of risk.

\subsection{Simulations}
This section provides further detail on the results provided in Section~\ref{sec:results-simulations}.
We used simulations to isolate the effects of the changes in the DMP due to the risk assessments. 
This required simulating outcomes in the four scenarios described in Table~\ref{tab:scenarios} and comparing the results of Scenario 3 and Scenario 4. 
We did this in two stages.
First, we used data from the experiment to learn participant prediction and decision functions both with and without the presence of a risk assessment. 
Second, we applied those functions to a large sample of defendants and loan applicants to simulate the outcomes of the four Table~\ref{tab:scenarios} scenarios.

\subsubsection{Fitting Prediction and Decision Models}
We began by learning the functions that explain the average risk prediction and negative decision rate for each subject. 
For predictions, we used Equations~\ref{eq:rpp_pretrial} and \ref{eq:rpp_loans}. 
For decisions, our formulas included the same factors as the predictions models, plus the perceived risk about that subject and the interaction between the perceived risk and whether the risk assessment was shown.
Equation~\ref{eq:sim_dmp_pretrial} is the regression in the pretrial setting and Equation~\ref{eq:sim_dmp_loans} is the regression in the loans setting.

\begin{equation}
\label{eq:sim_dmp_pretrial}
\begin{split}
\mathrm{detention.rate} 
&\sim \mathrm{perceived.risk + race + gender + age + offense.type} \\
&+ \mathrm{number.prior.arrests + number.prior.convictions} \\
&+ \mathrm{prior.failure.to.appear + show.RA + perceived.risk*show.RA} \\
&+ \mathrm{race*show.RA + gender*show.RA + age*show.RA} \\
&+ \mathrm{offense.type*show.RA + number.prior.arrests*show.RA} \\
&+ \mathrm{number.prior.convictions*show.RA + prior.failure.to.appear*show.RA}
\end{split}
\end{equation}

\begin{equation}
\label{eq:sim_dmp_loans}
\begin{split}
\mathrm{rejection.rate} 
&\sim \mathrm{perceived.risk + income + fico.category + own.home} \\
&+ \mathrm{monthly.installment + interest.rate + loan.amount + loan.term + show.RA} \\
&+ \mathrm{perceived.risk*show.RA + income*show.RA + fico.category*show.RA} \\
&+ \mathrm{own.home*show.RA + monthly.installment*show.RA} \\
&+ \mathrm{interest.rate*show.RA + loan.amount*show.RA + loan.term*show.RA}
\end{split}
\end{equation}

We fit all models using generalized linear regression with a logit link function from the quasibinomial family. 
We used this quasibinomial approach because the fitted values of these regressions are bounded probabilities (either risk predictions or negative decision rates, which both range from 0\%-100\%). 
Although linear regression yields very similar results, it does not guarantee that predicted values will be bounded between 0 and 1.

Before applying these models to new defendants, we used leave-one-out cross-validation to test the effectiveness of this approach on the data from our experiment. 
For each model in each setting, we removed one subject at a time, trained the model on the predictions or decisions about the other 299 subjects, and estimated the prediction or decision that would be made about the held-out subject both with and without the risk assessment. 
We evaluated these models by applying the risk prediction model and then using the output of that model as input to the decision model.
The mean average error (MAE) of the entire pipeline for negative decisions rates is 5.92 (RMSE=7.46) in the pretrial setting and 7.33 (RMSE=9.95) in the loans setting. 
All the models are unbiased estimators, with mean errors close to 0.

We then fit prediction and decision models for both settings on the complete set of 300 subjects for use in our simulations.

\subsubsection{Simulating Predictions and Decisions on New Subjects}
We applied these models to the held-out validation sets from both settings (not including the 300 subjects sampled from those datasets for inclusion in our experiment). 
These samples represent approximately 10\% of the complete data in each setting.
They contain 4,375 defendants and 4,231 loan applicants drawn from the populations described in Tables~\ref{tab:defendants} and \ref{tab:loans}. 

Our simulations proceeded as follows:
\begin{enumerate}
	\item Apply the predictions and decisions models to every subject to estimate the negative decision probabilities in the four scenarios from Table~\ref{tab:scenarios}. The predictions and decisions models enable us to simulate outcomes both with and without a risk assessment's advice. Using the outputs of the predictions model as the perceived risk, we applied these models in all four possible combinations of whether the risk assessment affected predictions and decisions. This process yields four estimated negative decision probabilities for each subject: predictions and decisions are both unaffected by the risk assessment (Scenario 1), predictions are unaffected by the risk assessment but decisions are affected by the risk assessment (Scenario 2), predictions are affected by the risk assessment but decisions are unaffected by the risk assessment (Scenario 3), and predictions and decisions are both affected by the risk assessment (Scenario 4).
	\item Run 1,000 trials simulating the outcome for each subject in each scenario, based on the negative decision probabilities found in the prior step. Doing this allowed us to estimate the distribution of outcomes for all four scenarios from Table~\ref{tab:scenarios}. 
\end{enumerate}

\section{Alternative Explanations}
\label{app:alternatives}
In this section, we discuss potential alternative explanations for our conclusion that showing the risk assessment altered the DMP and describe why they are inconsistent with our results. 

\subsection{Participants Have Greater Confidence in Risk Predictions}
One alternative explanation is that the risk assessment makes people more confident in their risk prediction rather than more concerned about avoiding risk in decision-making. 
In other words, people may place a greater weight on their risk prediction because they are more certain about this prediction rather than because they are more concerned about risk as a consideration. 
If this were the case, we would expect to see risk become a more ``extreme'' distinguishing factor in decisions: low levels of perceived risk lead to lower negative decision rates, while high levels of perceived risk lead to higher rates. 
Although that is what we observe in the pretrial setting, we observe a very different pattern in the loans setting: rejection rates go up at all levels of risk (Figure~\ref{fig:dmp_shifts}). 
The loans setting results are consistent with our explanation that the risk assessment makes people more risk-averse, yet inconsistent with people becoming more confident in their risk predictions. 
For instance, it is relatively implausible that becoming more confident that a loan applicant has a 0\% likelihood to default on the loan would more than double the likelihood of rejecting that loan application (Table~\ref{tab:decision_probs}). 

This pattern in the loans setting suggests that the pretrial setting results are also caused by greater attentiveness to risk rather than greater confidence in estimates of risk.
Furthermore, even if the pretrial setting does involve greater confidence in risk predictions, the effect would be equivalent to increasing the salience of risk: in both cases, the risk assessment would be causing perceived risk to become a stronger determinant of whether defendants are released or detained.

We can further investigate the role of confidence in decision-making by looking at participant self-reports of confidence. 
In the exit survey at the end of the experiment, we asked participants how confident they were in their decisions on a Likert scale from 1 (least confident) to 7 (most confident). 
We found that the risk assessment had no significant effects on participant confidence. 
In the pretrial setting, the risk assessment did not alter confidence among participants making predictions (P=.978, d=0.00) or decisions (P=.246, d=0.08).
Similarly, in the loans setting, the risk assessment did not alter confidence among participants making predictions (P=.580, d=0.07) or decisions (P=.213, d=0.09). 
Given that the risk assessments did not significantly impact participant self-reports of confidence, it is unlikely that the effects of the risk assessments can be attributed to them making participants more confident in their estimates of risk.

\subsection{Prediction-Makers and Decision-Makers Have Different Predictions of Risk}
Another alternative explanation is that perceived risk differs between participants making predictions and participants making decisions. 
In particular, the risk assessment might exert a stronger influence on participants making predictions than on participants making decisions.
Our results directly contradict this explanation, however.
Most notable is the contrast between the effects of the risk assessment in the loans setting, reducing predictions of risk without reducing loan rejections. 
For instance, among the 92.3\% of loan applicants for whom the risk assessment reduced perceived risk, almost half received a higher likelihood of rejection when the risk assessment was shown.
For this explanation to apply here, it would have to be the case that for almost half of the loan applicants, the risk assessment reduced risk estimates for prediction-makers yet increased risk estimates for decision-makers.
Although it is plausible that the risk assessment's effects on predictions could be attenuated for decision-makers, it is not plausible that prediction-makers and decision-makers would have their risk estimates influenced in opposite directions. 

\subsection{The Risk Assessment Provides a Random Shock to Decisions}
A third alternative explanation is that the risk assessments provide a random shock to decision-making, adding ``noise'' to decisions in a manner that is not connected to perceived risk. 
Two results clearly rule out this explanation. 
First, we observed that the reduction in pretrial detention was statistically significant, indicating that risk assessments can influence decisions in specific directions. 
Second, in both settings there was a positive and statistically significant relationship between changes in perceived risk and changes in negative decision rates for each subject (Figure~\ref{fig:shifts}).
These correlations indicate that the risk assessments' effect on decisions is (at least loosely) connected to the risk assessments' effect on perceived risk. 

\subsection{The Risk Assessment Alters the ``Other Factors'' Rather than the DMP}
Another potential explanation is that the risk assessment alters the calculation of the ``other factors'' that are incorporated into the DMP (Figure~\ref{fig:dmp}) rather than (or in addition to) altering the DMP itself. 
In the loans setting, for instance, the risk assessment could cause people to reduce their evaluation of the benefits of granting home improvement loans rather than cause people to become more risk-averse.
However, there is little reason to believe that receiving an algorithmic risk estimate would prompt a large enough reduction in perceived benefit to fully offset the large observed reductions in perceived risk. 
Moreover, although this alternative explanation would place the change at a different place in Figure~\ref{fig:dmp}, the overall effect would be similar: the risk assessment would be altering decision-making in unexpected ways that can have significant negative impacts.

\section{Tables}

\begin{table}[H]
\caption{Attributes of the full sample of defendants released before trial and the 300-defendant sample presented to participants in the experiment, by race. A violation means that the defendant was rearrested before trial, failed to appear for trial, or both.}
\begin{tabular}{@{}l|lll|lll@{}}
\toprule
 & \textbf{All} & \textbf{Black} & \textbf{White} & \textbf{Sample} & \textbf{Black} & \textbf{White} \\ 
 & N=47,141 & N=26,246 & N=20,895 & N=300 & N=189 & N=111 \\ \midrule
\textbf{Attributes} &  &  &  &  &  &  \\
Male & 76.7\% & 77.3\% & 75.5\% & 86.7\% & 88.4\% & 83.8\% \\
Black & 55.7\% & 100.0\% & 0.0\% & 63.0\% & 100.0\% & 0.0\% \\
Mean age at arrest & 30.8 & 30.1 & 31.8 & 28.1 & 27.1 & 29.8 \\
Drug crime & 36.9\% & 39.2\% & 34.0\% & 49.3\% & 50.8\% & 46.8\% \\
Property crime & 32.7\% & 30.7\% & 35.3\% & 30.3\% & 28.0\% & 34.2\% \\
Violent crime & 20.4\% & 20.9\% & 19.8\% & 14.0\% & 14.3\% & 13.5\% \\
Public order crime & 10.0\% & 9.3\% & 10.8\% & 6.3\% & 6.9\% & 5.4\% \\
Has prior arrest(s) & 63.4\% & 68.4\% & 57.0\% & 64.7\% & 73.5\% & 49.5\% \\
Mean number of prior arrests & 3.8 & 4.3 & 3.1 & 4.3 & 5.0 & 3.1 \\
Has prior conviction(s) & 46.5\% & 51.2\% & 40.7\% & 50.0\% & 57.7\% & 36.9\% \\
Mean number of prior convictions & 1.9 & 2.2 & 1.6 & 2.4 & 2.9 & 1.7 \\
Has prior failure(s) to appear & 25.1\% & 28.8 & 20.4\% & 31.7\% & 34.4\% & 27.0\% \\
 &  &  &  &  &  &  \\
\textbf{Outcome} & & & & & & \\
Rearrest & 15.0\% & 16.9\% & 12.6\% & 19.0\% & 20.1\% & 17.1\% \\
Failure to appear & 20.3\% & 22.6\% & 17.5\% & 25.3\% & 29.6\% & 18.0\% \\
Violation & 29.8\% & 33.1\% & 25.6\% & 36.0\% & 39.2\% & 30.6\% \\ \bottomrule
\end{tabular}
\label{tab:defendants}
\end{table}

\begin{table}[H]
\caption{Attributes of the full sample of approved home improvement loans and the 300-loan sample presented to participants in the experiment.}
\begin{tabular}{@{}lll@{}}
\toprule
 & \textbf{All} & \textbf{Sample} \\ 
 & N=45,218 & N=300 \\ \midrule
\textbf{Applicant} &  &  \\
Mean annual income & \$95,262.88 & \$93,349.22 \\
Mean credit score & 707.5 & 705.9 \\
Has a mortgage & 83.9\% & 83.0\% \\
 &  &  \\
\textbf{Loan} &  &  \\
Mean loan amount & \$14,556.38 & \$14,076.00 \\
Mean months to pay off loan & 42.4 & 42.6 \\
Mean monthly payment & \$435.75 & \$419.49 \\
Mean interest rate & 13.0\% & 13.2\% \\
 &  &  \\
\textbf{Outcome} &  &  \\
Loan paid off & 83.2\% & 84.7\% \\
Loan defaulted on & 16.8\% & 15.3\% \\ \bottomrule
\end{tabular}
\label{tab:loans}
\end{table}

\begin{table}[H]
\caption{Bayesian linear regression results estimating the risk-prediction process in both settings, following Equations~\ref{eq:rpp_pretrial} and \ref{eq:rpp_loans}. 
The first column presents the coefficient of each factor; the second column presents the coefficient of the interaction between that factor and the risk assessment being shown. 
The second column thus describes how showing the risk assessment altered each factor.
In the loans regression, annual income, loan amount, and monthly installment are measured in units of \$1,000. 
Parenthetical terms represent standard errors. . P<0.1; * P<0.05; ** P<0.01; *** P<0.001}
\begin{tabular}{@{}lll@{}}
\toprule
 & Not Shown RA & Shown RA (interaction) \\ \midrule
\textbf{Pretrial} &  &  \\
Intercept & 27.88 (1.50) *** & +6.98 (2.03) *** \\
White & –0.03 (0.72) & –0.98 (0.98) \\
Male & 0.04 (0.91) & –0.42 (1.25) \\
Age & 0.03 (0.04) & –0.20 (0.05) *** \\
Property crime & –2.29 (0.74) *** & +0.43 (1.04) \\
Public order crime & –0.28 (1.59) & –3.50 (2.21) \\
Violent crime & 3.00 (0.95) *** & –7.45 (1.27) *** \\
Number of prior arrests & 0.72 (0.17) *** & +0.20 (0.23) \\
Number of prior convictions & 0.31 (0.17) . & +0.09 (0.22) \\
Prior failure to appear & 27.82 (1.33) *** & –7.43 (1.76) *** \\
\textbf{} &  &  \\
\textbf{Loans} &  &  \\
Intercept & 39.37 (1.93) *** & –24.02 (2.45) *** \\
Annual income & –0.03 (0.01) *** & –0.02 (0.01) * \\
Good FICO score & –5.81 (1.04) *** & +2.23 (1.31) . \\
Very good FICO score & –7.91 (1.46) *** & +1.47 (1.83) \\
Exceptional FICO score & –9.29 (2.52) *** & –0.51 (3.24) \\
Fully own home & –0.30 (0.99) & +2.13 (1.26) . \\
Loan amount & 0.27 (0.28) & –0.45 (0.37) \\
Monthly installment & –0.74 (8.96) & +16.81 (11.50) \\
Interest rate & 0.33 (0.12) ** & +0.51 (0.15) *** \\
60-month term & –2.21 (1.91) & +7.41 (2.49) ** \\ \bottomrule
\end{tabular}
\label{tab:rpp_shifts}
\end{table}

\begin{table}[H]
\caption{Modeled probability of negative decisions at a range of perceived risk levels, by setting and risk assessment treatment. 
The negative decision in the pretrial setting is detaining the defendant; the negative decision in the loans setting is rejecting the loan application.
No RA indicates the probability of negative decisions when the risk assessment is not shown, Shown RA indicates the probability of negative decisions when the risk assessment is shown, and Difference indicates the difference between these values (numbers in brackets indicate the effect size of this difference). 
All differences in both settings are statistically significant with P<.001.
These results are plotted in Figure~\ref{fig:dmp_shifts}.}
\begin{tabular}{@{}l|lll|lll@{}}
\toprule
 & & \multicolumn{1}{c}{\textbf{Pretrial}} & & & \multicolumn{1}{c}{\textbf{Loans}} & \\
Perceived Risk & No RA & Shown RA & Difference & No RA & Shown RA & Difference \\  \midrule
0\% & 6.15\% & 2.06\% & –4.09\% {[}5.38{]} & 1.60\% & 3.24\% & +1.64\% {[}3.45{]} \\
10\% & 10.62\% & 4.74\% & –5.88\% {[}5.93{]} & 2.84\% & 5.98\% & +3.15\% {[}4.65{]} \\
20\% & 17.73\% & 10.52\% & –7.21\% {[}5.64{]} & 5.00\% & 10.82\% & +5.82\% {[}6.10{]} \\
30\% & 28.13\% & 21.80\% & –6.33\% {[}3.84{]} & 8.70\% & 18.81\% & +10.11\% {[}7.17{]} \\
40\% & 41.58\% & 39.83\% & –1.75\% {[}0.88{]} & 14.73\% & 30.68\% & +15.95\% {[}7.34{]} \\
50\% & 56.41\% & 61.11\% & +4.70\% {[}2.20{]} & 23.89\% & 45.78\% & +21.89\% {[}7.07{]} \\
60\% & 70.16\% & 78.83\% & +8.67\% {[}4.49{]} & 36.31\% & 61.63\% & +25.32\% {[}6.49{]} \\
70\% & 81.01\% & 89.80\% & +8.79\% {[}5.52{]} & 50.80\% & 75.27\% & +24.47\% {[}5.39{]} \\
80\% & 88.54\% & 95.41\% & +6.87\% {[}5.35{]} & 65.04\% & 85.19\% & +20.14\% {[}4.14{]} \\
90\% & 93.32\% & 98.00\% & +4.68\% {[}4.71{]} & 76.93\% & 91.55\% & +14.63\% {[}3.19{]} \\
100\% & 96.19\% & 99.14\% & +2.95\% {[}4.07{]} & 85.60\% & 95.32\% & +9.72\% {[}2.54{]} \\ \bottomrule
\end{tabular}
\label{tab:decision_probs}
\end{table}

\begin{table}[H]
\caption{Participant beliefs about how decision-makers should balance priorities. 
After making decisions, participants were asked to what extent a decision-maker (i.e., a judge or government loan agent) should value four salient considerations when making decisions. 
Participants had to assign a total of 100 points (in increments of 5) across the four considerations. 
None of the average values assigned to these considerations differ significantly across the control (Not Shown RA) and treatment (Shown RA) groups.}
\begin{tabular}{@{}lllll@{}}
\toprule
 & Not Shown RA & Shown RA & P-value & Effect size \\ \midrule
\textbf{Pretrial} &  &  &  &  \\
Incapacitation & 30.86 & 29.89 & .341 & 0.07 \\
Freedom & 25.76 & 26.68 & .372 & 0.07 \\
Deterrence & 20.04 & 19.05 & .245 & 0.08 \\
Rehabilitation & 23.35 & 24.38 & .289 & 0.08 \\
 &  &  &  &  \\
\textbf{Loans} &  &  &  &  \\
Likelihood to pay & 40.98 & 39.28 & .211 & 0.09 \\
Equity & 21.51 & 22.59 & .124 & 0.11 \\
Economic development & 19.63 & 19.29 & .622 & 0.03 \\
Neighborhood stability & 17.89 & 18.84 & .200 & 0.09 \\ \bottomrule
\end{tabular}
\label{tab:priorities}
\end{table}

\end{document}